\documentclass[preprint,showpacs,aps]{revtex4}

\usepackage{epsfig}

\begin{document}
\title{\large \bf Depinning phase transition in two-dimensional clock model with quenched randomness}

\author{X.P. Qin$^{1,2}$, B. Zheng$^{1}$\footnote{corresponding author; email: zheng@zimp.zju.edu.cn} and N.J. Zhou$^{3}$}

\affiliation{$^1$ Department of Physics, Zhejiang University, Hangzhou 310027, P.R. China\\
$^2$ School of Mathematics, Physics and Information Science,
Zhejiang Ocean University, Zhoushan 316000, P.R. China\\
$^3$ Department of Physics, Hangzhou Normal University, Hangzhou 310036, P.R. China}

\begin{abstract}
With Monte Carlo simulations, we systematically investigate the depinning phase transition in the two-dimensional
driven random-field clock model. Based on the short-time dynamic approach,
we determine the transition field and
critical exponents. The results show that
the critical exponents vary with the form of the random-field distribution and the strength of the
random fields, and the roughening dynamics of the domain interface
belongs to the new subclass with $\zeta \neq \zeta_{loc} \neq \zeta_s$ and $\zeta_{loc} \neq 1$.
More importantly, we find that the transition field and critical exponents change with
the initial orientations of the magnetization of the two ordered domains.
\end{abstract}


\maketitle

\section{Introduction}

In recent years the domain-wall dynamics in disordered media has
attracted much attention, which is important from both experimental and theoretical
perspectives. A crucial feature of the domain-wall motion in
disordered media is the depinning phase transition at zero
temperature. A typical theoretical approach to the domain-wall dynamics is
the Edwards-Wilkinson equation with quenched disorder (QEW)
\cite{edw82,nat92,ros03,due05,kol06,bus08}.
It is usually believed that different variants of the model
belong to a same universality class.

The QEW equation is a phenomenological model, without microscopic
structures and interactions of the materials. In two dimensions, the
domain wall is assumed to be a single-valued elastic string.
An alternative, possibly more realistic approach, is to construct
lattice models at the microscopic level.
The simplest example is the
two-dimensional ($2D$) driven random-field Ising model
(DRFIM) \cite{now98,rot99,rot01,zho09,zho10}. In the DRFIM model,
overhangs and islands are created in the dynamic evolution,
and the domain wall is not single-valued.
Very recently it has been demonstrated that the DRFIM model does not belong to the universality
class of the QEW equation, due to the dynamic effect of overhangs and islands \cite{zho09,zho10,don12},
and its results are closer to experiments \cite{met07,lem98,jos98,zho10a}.
The DRFIM model does not suffer from the theoretical self-inconsistence
as in the QEW equation, and may go beyond the QEW equation, for example, to describe the relaxation-to-creep transition \cite{ros01,zho10a}.
To distinguish the DRFIM model and QEW equation, one needs accurate measurements of the critical exponents.
In this respect, the short-time dynamic approach has been proven to be efficient \cite{kol06,zhe98,luo98,rod07,kol09,zho09,zho10}.

Traditionally domain walls in magnetic materials are viewed as Ising-like \cite{cat12}.
However, this assumption has been challenged by the complex Bloch walls or N\'eel walls. In recent years many activities have been devoted
to the orientation dynamics of the magnetization distribution, in experimental, theoretical and numerical studies
\cite{rhe10,uhl10,gou10,mar07,shi07}. Particular attention is also drawn to the depinning transition \cite{luo07}.
Recently domain walls converging at different angles are observed in the thin films of BiFeO$_3$, PbTiO$_3$ and PbZr$_{0.2}$Ti$_{0.8}$O$_3$ \cite{nel11,jia11}.

Up to now, most theoretical studies of the depinning phase transition of the domain interface
in magnetic materials are concentrated on the Ising-like systems, such as
the DRFIM model. In both theories and experiments, it is desirable to go beyond.
For example, one may consider the $XY$ model, in which
the spins are planar unit vectors. However, the pinning phenomenon does not
exist in the $2D$ $XY$ model. Possibly, this might be connected to the fact that
there are no real ordered states at finite temperatures in the $2D$ $XY$ model.

The $p$-state clock model is a
discrete version of the \emph{XY} model \cite{cha86,cze96}. In the equilibrium state,
the $2$-state clock model is just the Ising model; for $p=4$, the clock model is decoupled into two Ising models; for $p=3$,
it is equivalent to the three-state Potts model; for
$p\rightarrow\infty$, the clock model approaches the \emph{XY} model \cite{bae09}. For
$4<p<\infty$, there is a quasiliquid intermediate phase between two
transition temperatures $T_1$ and $T_2$ ($T_1<T_2)$. Here $T_1$ is
the transition temperature from the ordered phase to the quasiliquid phase, and
$T_2$ is that from the quasiliquid phase to the disordered
phase \cite{lap06,bri07,bri10}.

With Monte Carlo simulations, and based on the short-time dynamic approach, the purpose of this paper is to study the depinning phase transition in the
$2D$ $p$-state clock model with quenched random fields, emphasizing the dynamic effect of the orientations of the magnetization of the two ordered domains.
Numerical results are mainly presented for $p=6$, and also extended from $p=2$ to $12$.
In Sec. II, the model and scaling
analysis are described. In Sec. III, numerical results are
presented. Sec. IV includes the conclusions.

\section{Model and scaling analysis}

\subsection{Model}
The $2D$ $p$-state driven random-field clock model (DRFCM) is defined on a square lattice by the
Hamiltonian
\begin{eqnarray}
\mathcal{H} &=& - J \sum_{<ij>}\vec{S_i}\cdot\vec{S_j} - \sum_i (\vec H+\vec{h_i}) \cdot \vec S_{i} \nonumber \\
&=& - J \sum_{<ij>}\cos(\theta_i - \theta_j) - H \sum_i \cos\theta_i
- \sum_i h_i \cos(\varphi_i-\theta_i), \label{equ10}
\end{eqnarray}
where the spin $\vec{S}_{i}=(\cos\theta_i,\sin\theta_i)=(S_{i,x},S_{i,y})$ is
a planar unit vector at site $i$ of a rectangle $L_x\times
L_y$, and $\theta_i$ is the angle associated with the orientation of the
spin, taking discrete
values $\theta_i=2\pi q_i/p$ with $q_i=0,\ldots,p-1$. The sum $<ij>$ is
over the nearest neighbors, $\vec H$ is a homogeneous external driving field in the $x$ direction,
 and $\varphi_i$ is the angle associated with the orientation of  the random-field $\vec{h_i}$, taking discrete values $\varphi_i=2\pi
n_i/p$. Here $n_i$ is a random integer within an interval
$[0,p-1]$. The random field $h_i$ may be distributed in different forms.
A typical example is a uniform distribution within an interval
$[-\Delta,\Delta]$, following Refs.~\cite{now98,zho09,zho10}.
We take $\Delta$ from
$1.0J$ to $2.5J$ and set $J=1$.
Another example is a Gaussian distribution with standard deviation
$\sigma$ and zero mean. The form of the random-field distribution and the strength of the
random fields may change
the order of the depinning phase transitions and also its universality class \cite{ji91,koi10}.

We consider a dynamic process with a semiordered initial
state with a perfect domain wall in the $y$ direction.
As time evolves, the domain wall propagates and roughens,
therefore it looks like an interface. We first concentrate on the uniform distribution of the random fields,
and then extend the calculations to the Gaussian distribution.
Our simulations are performed at zero temperature with
lattice sizes $L_y=256, 512, 1024$, and $2048$ up to $t_{max}=5120$.
Total samples for average are about 20000 for different random
fields. Main results are presented with $L_y=1024$ and
simulations of different $L_y$ confirm that finite-size effects are
already negligibly small. $L_x$ is taken to be sufficiently large so
that the boundary is not reached. Typically, $L_x=300$ to $500$. Statistical errors are estimated
by dividing the total samples in a few subgroups. If the
fluctuation in the time direction is comparable with or larger than
the statistical error, it will be taken into account.

For an even $p$, spins in the initial state take the orientation $\theta_L=0$ on the sublattice at
the left side, while $\theta_R=\pi$ on the
sublattice at the right side. For an odd $p$, spins in the initial state take the orientation $\theta_L=0$ on the sublattice at
the left side, while $\theta_R= \pi (p-1)/p$ on the
sublattice at the right side.
Further, we also change the initial orientation $\theta_R$ on the
sublattice at the right side, to examine its influence on the depinning phase transition.
A periodic boundary condition is used in the $y$
direction, while spins at the boundary in the $x$ direction are
fixed. To eliminate the pinning effect irrelevant for the disorder, we
rotate the lattice such that the initial domain wall orients in the
$(11)$ direction of the lattice \cite{now98,zho09}.

\begin{figure}[ht]
\epsfysize=7.cm \epsfclipoff \fboxsep=0pt
\setlength{\unitlength}{1.cm}
\begin{picture}(10,7)(0,0)
\put(-3.2,0.){{\epsffile{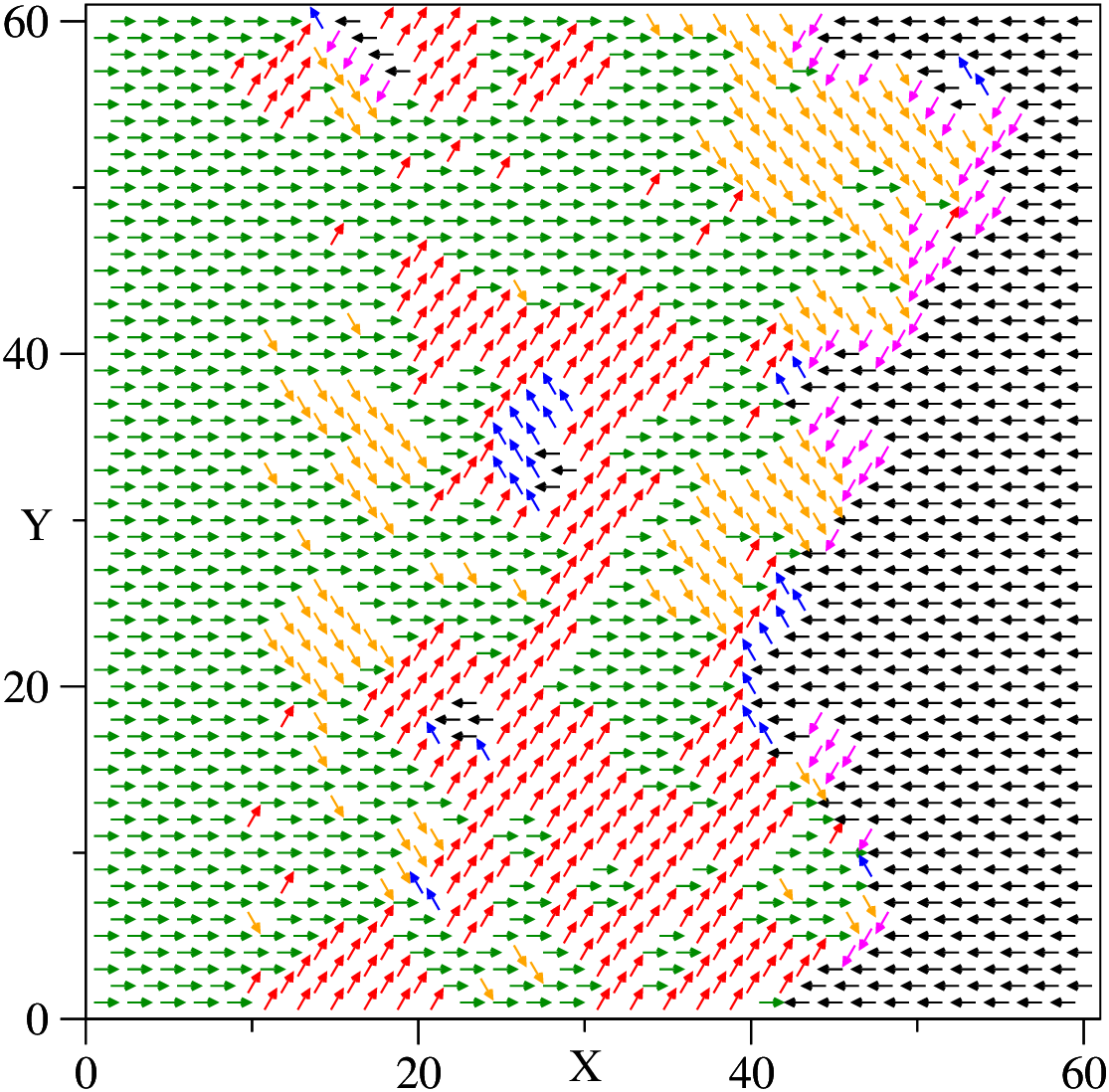}}}\epsfysize=7.cm
\put(5.3,0.){{\epsffile{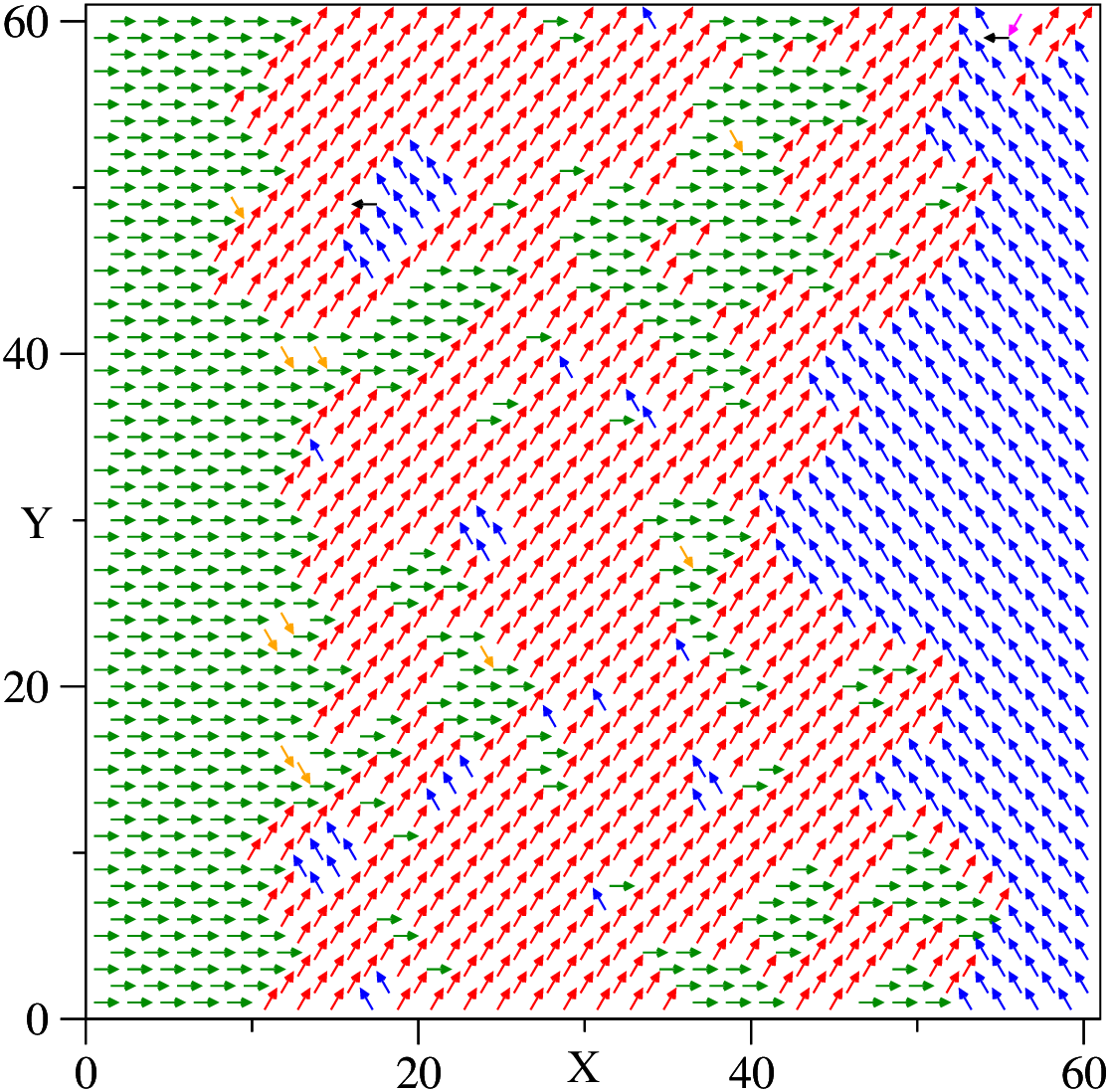}}}
\put(-2.5,-0.4){\footnotesize(a)~$\sigma=1.5, H=0.9643$}
\put(5.9,-0.4){\footnotesize(b)~$\sigma=1.5, H=0.8020$}
\end{picture}
\caption{(Color online)Snapshots of the spin
configurations at $t=500$ Monte Carlo time
steps, for the $6$-state DRFCM model with the Gaussian distribution of the random fields at $H_c$.
Arrowheads denote the orientation of the spins. (a) $\theta_L=0$ and $\theta_R=\pi$; (b) $\theta_L=0$ and $\theta_R=2\pi/3$.}
\label{evolution}
\end{figure}

After preparing the initial state, we
randomly select a spin and a new state $q_i$, and
flip it if the total energy decreases after flipping,
and at least one of the nearest neighbors is already flipped \cite{cha86,zho09}. In other words, only spins at the interface are allowed to be
flipped \cite{dro98,koi10}. A Monte Carlo time step is defined as $L_x\times
L_y$ single spin flips. In Fig.~\ref{evolution}, snapshots of the spin configurations are displayed for the $6$-state DRFCM model with
the Gaussian distribution of the random fields, to show the roughening of the domain interface.
Overhangs and islands are created.
In particular, due to the existence of the intermediate states of the spin $\vec S_i$,
intermediate regimes emerge. The microscopic structure of the interface becomes more complicated
than that of the DRFIM model \cite{zho09,zho10}.

There may be different ways to define the domain interface. We adopt the
definition with the magnetization. Other reasonable definitions yield similar results. Denoting a spin at site $(x,y)$
by $\vec{S}_{xy}(t)=(\cos\theta_{xy}(t),\sin\theta_{xy}(t))$, we introduce a
line magnetization of the $x$ component \cite{cze96,bri07,bri10}
\begin{equation}
m(y,t) = \frac{1}{L_x} \left[ \sum_{x=1}^{L_x} \cos\theta_{xy}(t)
\right]. \label{equ20}
\end{equation}
The height function of the domain interface is then defined as
\begin{equation}
h(y,t) = \frac{L_x}{2}[ m(y,t) + 1] . \label{equ30}
\end{equation}
Following Refs.~\cite{now98,zho09,zho10,ram00,che10}, to describe the propagating and roughening of the domain interface,
we define the average velocity $v(t)$,
roughness function $\omega^2(t)$, and height
correlation function $C(r, t)$ of the domain interface respectively,
\begin{equation}
v(t) = \langle \frac{dh(y,t)}{dt} \rangle , \label{equ40}
\end{equation}
\begin{equation}
\omega^2(t) = \left \langle h(y,t)^2 \right \rangle - \langle h(y,t)
\rangle^2, \label{equ50}
\end{equation}
\begin{equation}
C(r, t) = \left\langle[h(y + r, t) - h(y, t)]^2 \right \rangle.
\label{equ60}
\end{equation}
Here $\langle \cdots \rangle$ represents not only the statistic average but also
the average in the $y$ direction.
Further, we introduce the Fourier transform of the height correlation function,
\begin{equation}
h(k,t) = \frac{1}{\sqrt{L_y}} \sum_{y=1}^{L_y} h(y,t)exp(iky), \label{equ70}
\end{equation}
and define the structure factor
\begin{equation}
S(k,t) = \langle h(k,t)h(-k,t) \rangle . \label{equ80}
\end{equation}
In Eq.~(\ref{equ80}), $\langle \cdots \rangle$ represents only the statistical
average.

To independently estimate the dynamic exponent $z$, we introduce a quantity
\begin{equation}
F(t) = [M^{(2)}(t) - M(t)^2]/\omega^2(t). \label{equ90}
\end{equation}
Here $M(t)$ is the global magnetization and $M^{(2)}(t)$ is
its second moment. In fact, $F(t)$ is nothing but the ratio of the
planar susceptibility and line susceptibility.

\subsection{Scaling analysis}

For the uniform distribution with $\Delta > \Delta_c$
and Gaussian distribution of the random fields,
the depinning phase transition is of second-order.
Here $\Delta_c$ is the critical strength of the random
fields separating the first-order phase transition and second-order one.
For the depinning transition, the order parameter is the average velocity $v(t)$ of the domain interface, and it should obey the dynamic
scaling theory supported by the renormalization-group calculations
\cite{jan89,zhe98,luo98}. In the critical regime, there are two
spatial length scales in the dynamic relaxation, i.e., the
nonequilibrium spatial correlation $\xi(t)\sim t^{1/z}$ and the
lattice size $L_y$, and scaling arguments lead to the scaling
form \cite{jan89,zhe98,luo98},
\begin{equation}
v(t, \tau, L_y) = b^{-\beta/\nu} G(b^{-z}t, b^{1/\nu}\tau,
b^{-1}L_y). \label{equ100}
\end{equation}
Here, $\tau=(H-H_c)/H_c$, $H_c$ is the depinning transition field, $b$ is an arbitrary rescaling factor, $z$ is the dynamic
exponent, $\beta$ and $\nu$ are the static exponents.
This dynamic scaling form is assumed to hold already in the macroscopic short-time regime, after a microscopic time scale $t_{mic}$,
which is sufficiently large in the microscopic sense.
Taking $b \sim t^{1/z}$, the dynamic scaling
form is rewritten as
\begin{equation}
v(t, \tau, L_y) = t^{-\beta/\nu z} G(1, t^{1/\nu z}\tau, t^{-1/z}L_y).
\label{equ110}
\end{equation}
In the short-time regime, i.e., the regime with $\xi(t) \ll L_y$,
the finite-size effect is negligible,
\begin{equation}
v(t, \tau) = t^{-\beta / \nu z}G(t^{1/\nu z}\tau). \label{equ120}
\end{equation}
At the transition field $H_c$, i.e., $\tau = 0$, a power-law behavior
is achieved,
\begin{equation}
v(t) \sim t^{-\beta / \nu z}. \label{equ130}
\end{equation}
With Eq.~(\ref{equ120}), one may locate the transition
field $H_c$ by searching for the best power-law behavior of the interface velocity $v(t, \tau)$
\cite{zhe98,luo98}. To determine $\nu$, one derives from
Eq.~(\ref{equ120})
\begin{equation}
\partial _\tau \ln v(t,\tau)|_{\tau=0} \sim t^{1 / \nu z}. \label{equ140}
\end{equation}
In general, even at the transition point, i.e., $\tau=0$, the
roughness function $\omega^2(t)$ and height correlation function
$C(r,t)$ may not obey a perfect power-law behavior in early times.
Due to the random updating scheme in numerical simulations, the
domain wall also roughens even without quenched disorder
($\Delta=0$, or $\sigma=0$). This leads to corrections to scaling. To capture the
dynamic effects of the disorder, therefore, we introduce the pure roughness
function and pure height correlation function respectively,
\begin{equation}
D\omega^2(t) = \omega^2(t) - \omega^2_b(t), \label{equ150}
\end{equation}
\begin{equation}
DC(r,t) = C(r,t)- C_b(r,t), \label{equ160}
\end{equation}
where $\omega^2_b(t)$ and $C_b(r,t)$ are the roughness function and
height correlation function without disorder. At
the transition point $\tau=0$ and for a sufficiently large lattice $L_y$,
$D\omega^2(t)$ and $DC(r,t)$ should obey the power-law scaling forms
\cite{jos96,zho08,bak08,zho09},
\begin{equation}
D\omega^2(t) \sim t^{ 2 \zeta / z}, \label{equ170}
\end{equation}
and
\begin{equation}
   DC(r,t) \sim  \left\{
   \begin{array}{lll}
     t^{2(\zeta-\zeta_{loc})/z}\ r^{2\zeta_{loc}}   & \quad &  \mbox{if~  $r \ll \xi(t)$} \\
     t^{2\zeta /z}  & \quad &  \mbox{if~ $ \xi(t) \ll r$}
   \end{array}\right. .
   \label{equ180}
\end{equation}
Here $\xi(t) \sim t^{1/z}$, $\zeta$ is the global roughness
exponent, and $\zeta_{loc}$ is the local roughness exponent.

The structure factor generally scales as \cite{ram00,che10},
\begin{equation}
S(k,t) = k^{-(2 \zeta +1)} f_s(kt^{1/z}), \label{equ190}
\end{equation}
where the scaling function takes the form,
\begin{equation}
   f_s(u) \sim  \left\{
   \begin{array}{lll}
     u^{2(\zeta-\zeta_s)}  & \quad &  \mbox{if ~ $u \gg 1 $} \\
     u^{2\zeta+1}  & \quad &  \mbox{if ~ $u \ll 1$}
   \end{array}\right. ,
   \label{equ200}
\end{equation}
and $\zeta_s$ is the spectral roughness exponent.

Since $\omega^2(t)$ represents the fluctuation in the $x$
direction and $M^{(2)}(t) - M(t)^2$ includes those in both the $x$
and $y$ directions, the dynamic scaling behavior of $F(t)$
should be
\begin{equation}
F(t) \sim \xi(t)/L_y \sim t^{1/z}/L_y. \label{equ210}
\end{equation}

\section{Monte Carlo Simulation}

Our main results are presented for the $6$-state DRFCM model, and those for other $p$ are also discussed.

\subsection{Uniform distribution of random fields}

\begin{figure}[ht]
\epsfysize=7.1cm \epsfclipoff \fboxsep=0pt
\setlength{\unitlength}{1.cm}
\begin{picture}(10,7)(0,0)
\put(-3.2,0.){{\epsffile{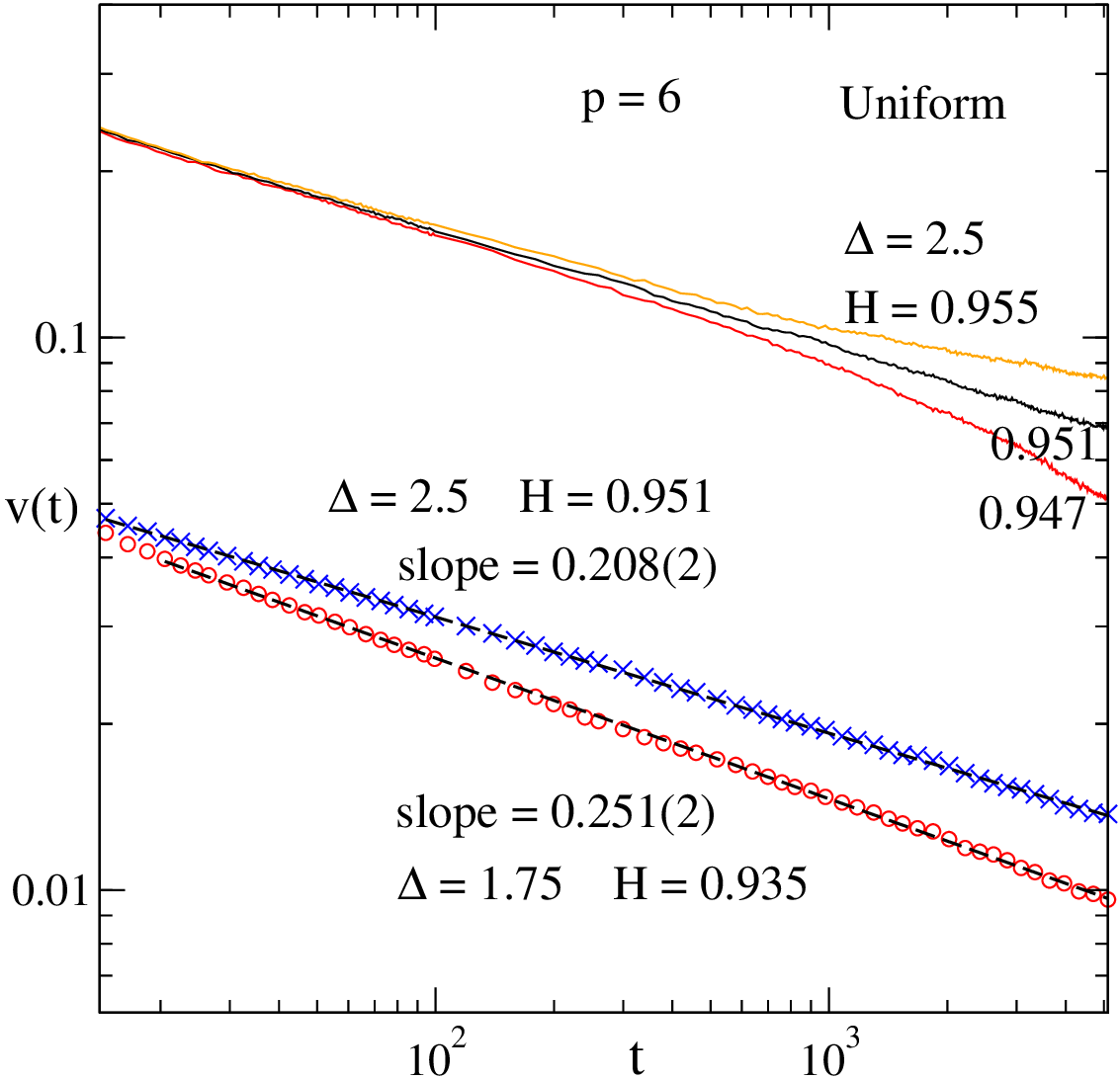}}}\epsfysize=7.1cm
\put(5.3,0.){{\epsffile{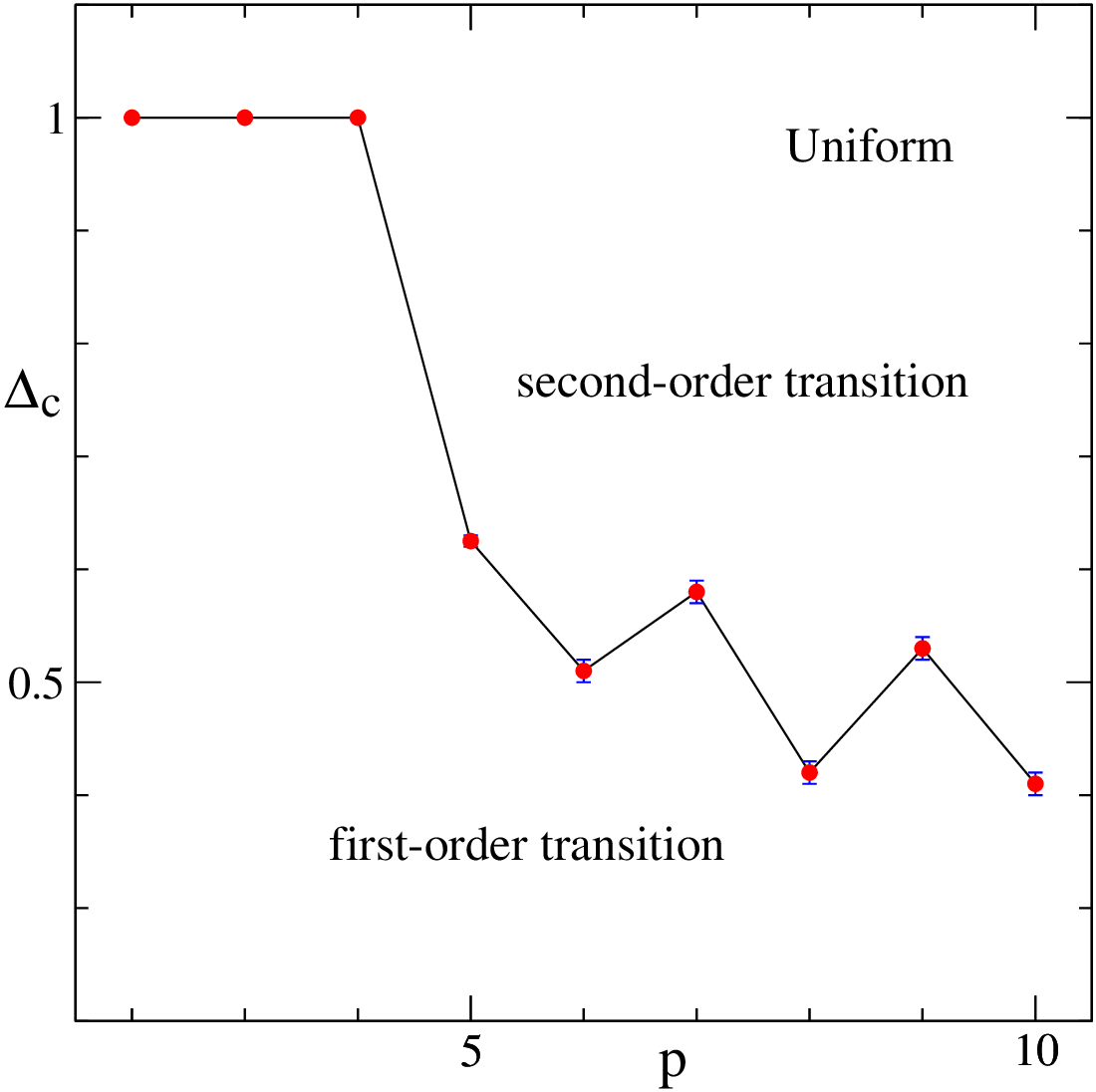}}}
\put(-2.2,-0.4){\footnotesize{(a)}}\put(6.2,-0.4){\footnotesize{(b)}}
\end{picture}
\caption{(Color online)(a) The interface velocity $v(t)$ is plotted for $\Delta=2.5$ at different
driving field $H$ with solid lines. For comparison,
$v(t)$ is also shown for $\Delta=1.75$ and $2.5$ at $H_c=0.935$
and $0.951$ respectively with circles and crosses. Dashed lines show
power-law fits. For clarity, the solid lines are shifted up by a factor of five. (b) The
critical strength $\Delta_c$ of the disorder is displayed for different $p$.
In both (a) and (b), the results are for the uniform distribution of the random fields.}\label{vt}
\end{figure}

We first study the depinning phase transition with the uniform distribution
of the random fields. In Fig.~\ref{vt}(a), the interface
velocity $v(t)$ is displayed for
different strengths of the random fields and different driving fields. For example, at $\Delta=2.5$,
it drops rapidly for a smaller $H$, while approaches a constant for
a larger $H$. By searching for the best power-law behavior, one
locates the transition field $H_c=0.935(1)$ and $0.951(1)$ for
$\Delta = 1.75$ and $2.5$, respectively.
Especially, the microscopic time scale $t_{mic}$ here is very small, about $10$ Monte Carlo time steps.
According to
Eq.~(\ref{equ130}), one measures the exponent $\beta /\nu
z=0.251(2)$ and $0.208(2)$ from the slopes of the curves for $\Delta
= 1.75$ and $2.5$, respectively.
The power-law behavior in Fig.~\ref{vt}(a)
indicates that the depinning transition is of second order.

For a smaller $\Delta$, the velocity approaches a constant for $H \geq
\Delta$, while drops rapidly to zero once $H$ is below $\Delta$. The power-law behavior in Eq.~(\ref{equ130}) could not be detected.
This is a signal that the depinning transition is of
first order, and the transition field
$H_c=\Delta$. The critical strength $\Delta_c$ of the random fields, separating the first-order depinning transition and second-order one,
varies with the state number $p$ of the spin $\vec S_i$, as shown in Table~\ref{t3} and Fig.~\ref{vt}(b).
In fact, this transition at $\Delta_c$ could be also detected from the change of the morphology of the domain interface \cite{ji91,koi10}.
For the $2D$ DRFIM model, as $\Delta$ increases, the morphology of the domain interface changes from faceted to self-similar at $\Delta_c=1$.

\begin{figure}[ht]
\epsfysize=7.1cm \epsfclipoff \fboxsep=0pt
\setlength{\unitlength}{1.cm}
\begin{picture}(10,7)(0,0)
\put(-3.2,-0.3){{\epsffile{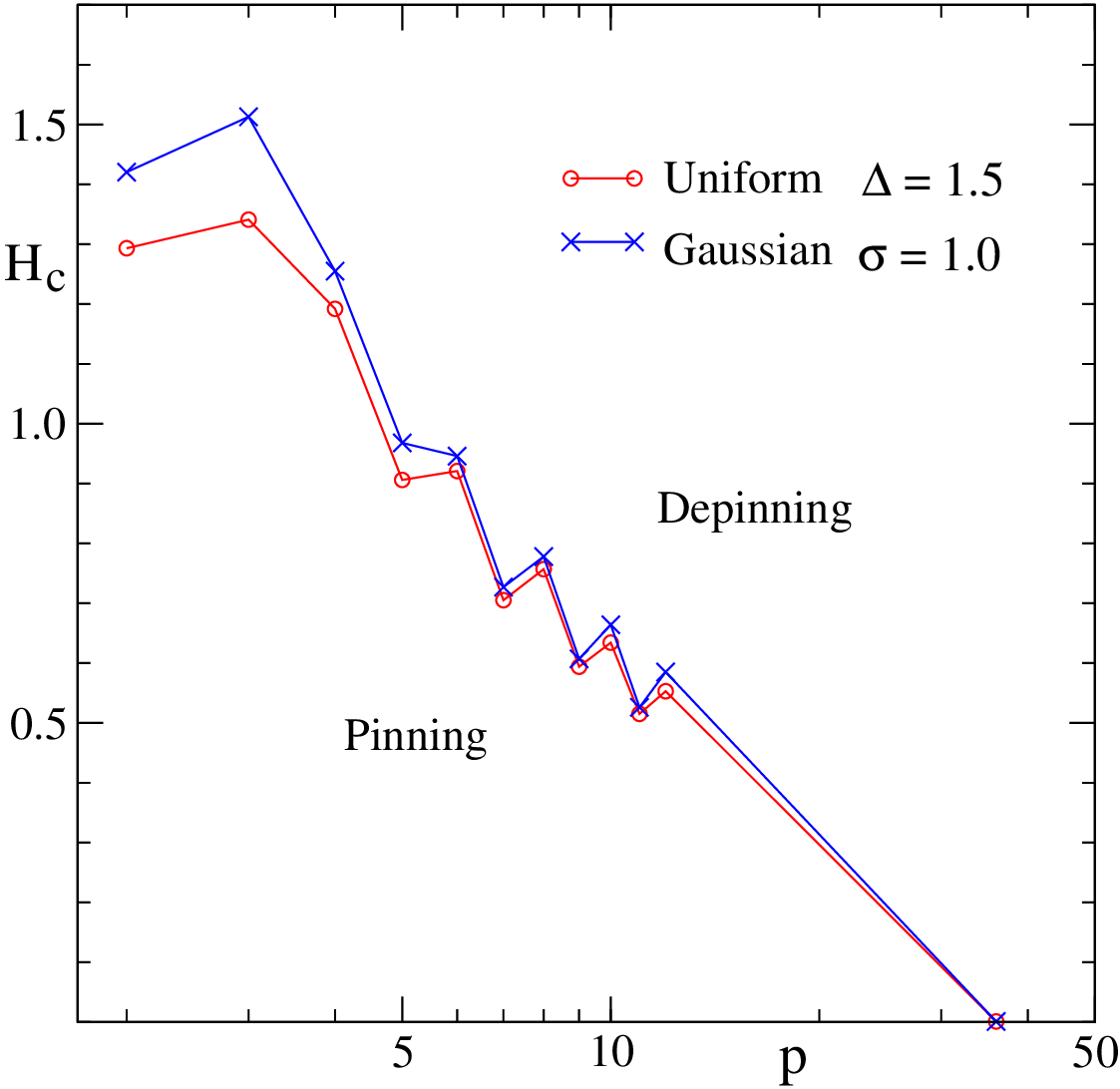}}}\epsfysize=7.1cm
\put(5.3,-0.3){{\epsffile{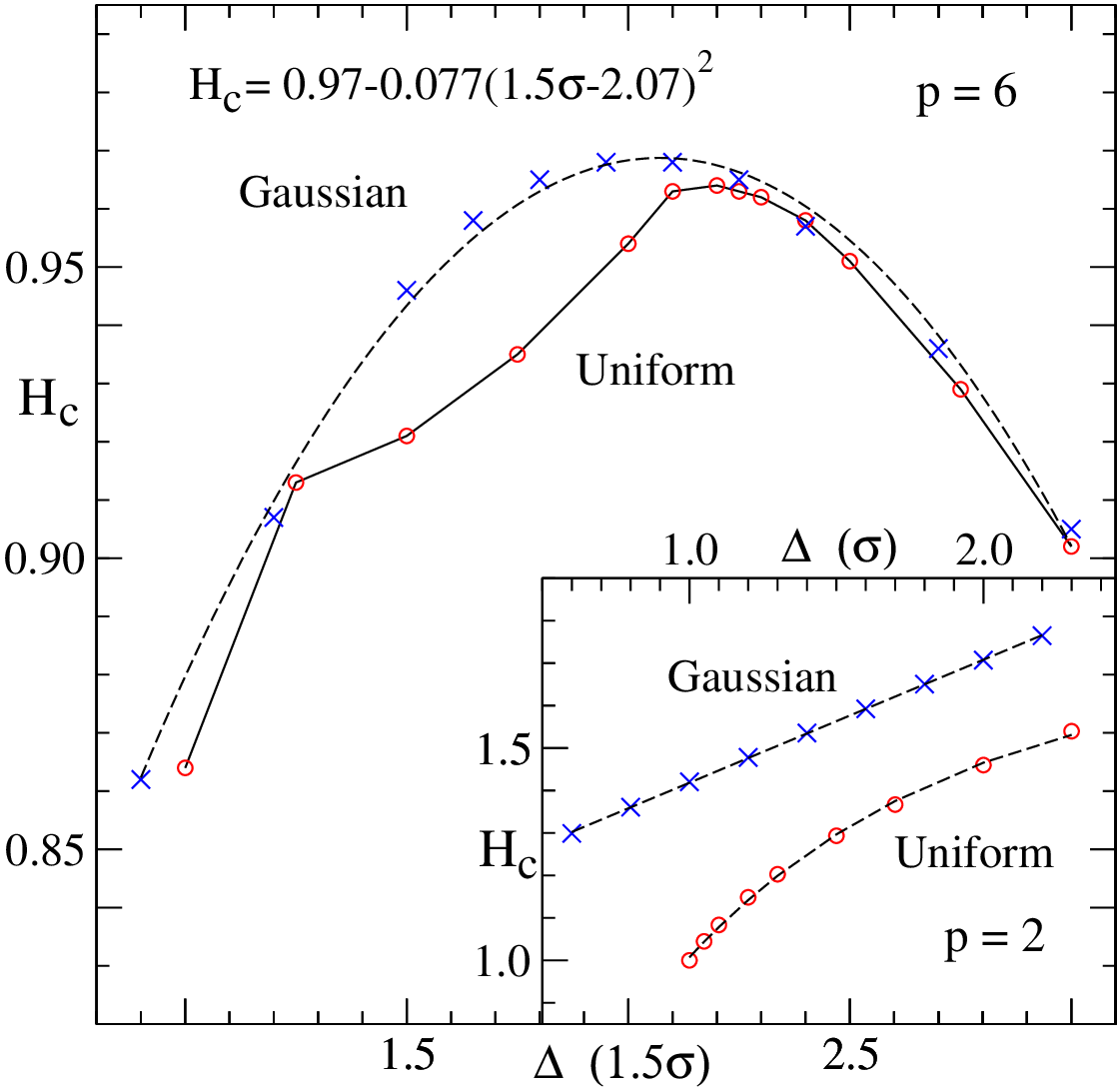}}}
\put(-2.2,-0.4){\footnotesize{(a)}}\put(6.2,-0.4){\footnotesize{(b)}}
\end{picture}
\caption{(Color online)(a) The depinning transition field $H_c$ is displayed for different $p$
at $\Delta=1.5$ and $\sigma=1.0$ for the uniform and Gaussian
distributions of the random fields respectively.
Errors are smaller than the symbol sizes. (b) $H_c$ is plotted for different strengths of the random fields. In the inset, $H_c$ for $p=2$, i.e., the DRFIM model, is shown for comparison.}\label{Hc}
\end{figure}

The transition field $H_c$ varies with $p$ and
$\Delta$. In Fig.~\ref{Hc}(a), $H_c$ is displayed for
different $p$ at $\Delta=1.5$, and the results
are also listed in Table~\ref{t3}. For $p=2$, i.e., the
DRFIM model, the transition field
$H_c=1.2933$ agrees with that in Ref.~\cite{zho09}. For a large $p$, e.g., $p=36$,
$H_c$ becomes very small, $H_c<0.001$. One expects that
$H_c \to 0$ in the limit $p \to \infty$. This is consistent with the fact that the depinning transition does not exist
in the $2D$ $XY$ model with quenched random fields.
The $p$-dependence of $H_c$ is different for even and odd $p$ because of the asymmetric initial state for odd $p$.
A similar behavior is also observed for the $p$-dependence of $\Delta_c$ in Fig.~\ref{vt}(b).
We will come back to this point in Subsec. III C.
In Fig.~\ref{Hc}(b), the transition fields $H_c$ is plotted for different strengths of the random fields,
and parts of the results are also given in Table~\ref{t1}. Different from the case of $p=2$,
$H_c$ of the $6$-state DRFCM model exhibits a non-monotonous and unusual behavior with respect to the strength of the random fields.
At $\Delta = 2.2$, $H_c$ reaches the maximum.

\begin{figure}[ht]
\epsfysize=7.1cm \epsfclipoff \fboxsep=0pt
\setlength{\unitlength}{1.cm}
\begin{picture}(10,7)(0,0)
\put(-3.2,-0.3){{\epsffile{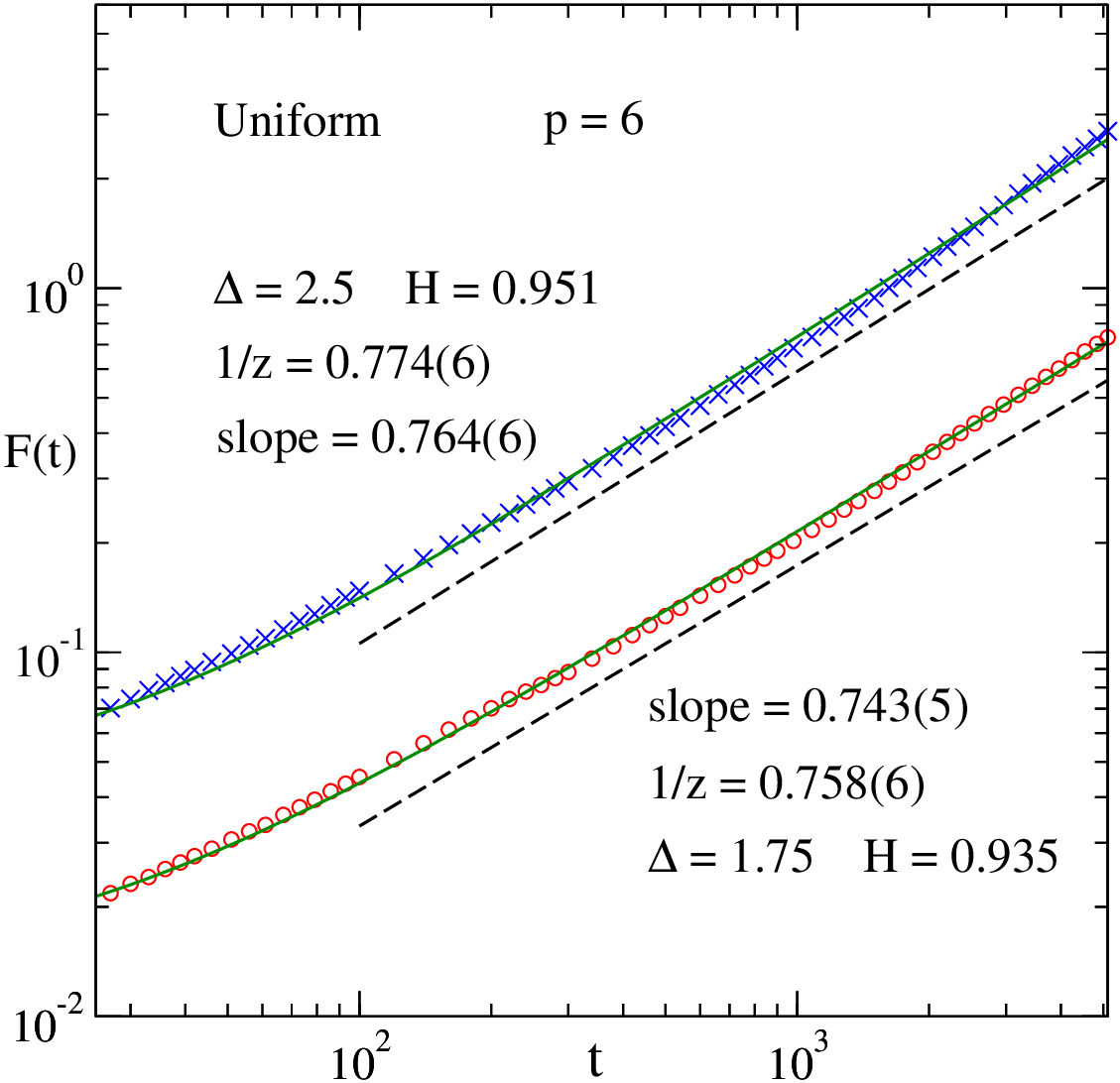}}}\epsfysize=7.1cm
\put(5.2,-0.3){{\epsffile{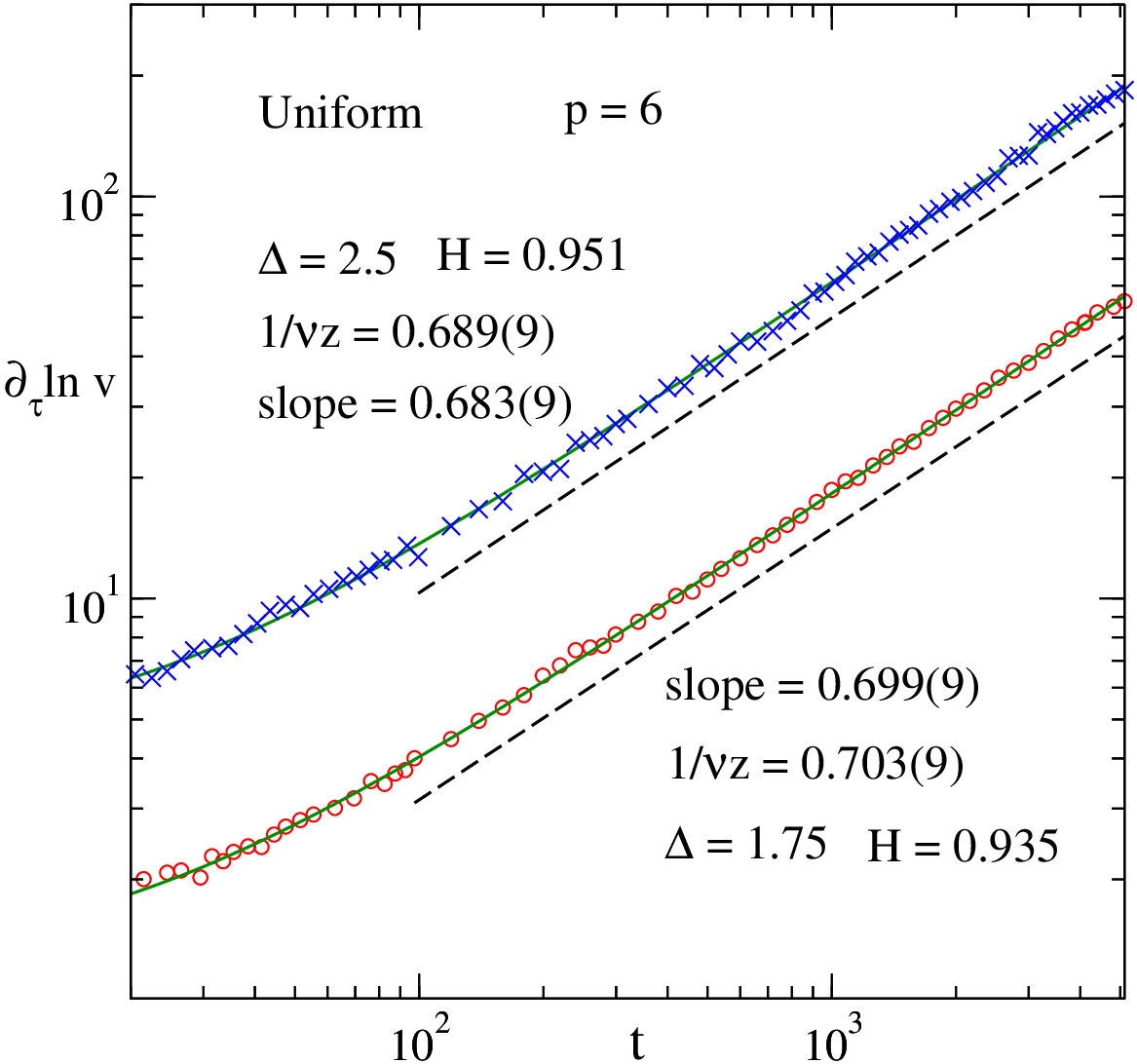}}}
\put(-2.2,-0.4){\footnotesize{(a)}}\put(6.5,-0.4){\footnotesize{(b)}}
\end{picture}
\caption{(Color online)(a) The quantity $F(t)$ is shown for $\Delta = 1.75$ and $2.5$ at
$H_c$. (b) The logarithmic derivative of $v(t,\tau)$ is plotted for
$\Delta = 1.75$ and $2.5$ at $H_c$. In both (a) and (b),
the results are for the uniform distribution of the random fields.
Dashed lines show power-law fits, and solid
lines include corrections to scaling. For clarity,
the curves for $\Delta = 2.5$ are shifted up by a factor of three. }\label{Ft}
\end{figure}

In Fig.~\ref{Ft}(a), the quantity $F(t)$ defined in
Eq.~(\ref{equ90}) is shown for $\Delta = 1.75$ and $2.5$ at $H_c$. A power-law behavior is
detected but with certain corrections to scaling up to a microscopic time scale $t_{mic} \sim 100$ or $200$, probably due to the multiple states of the spin $\vec S_i$.
According to Eq.~(\ref{equ210}), a direct measurement from the slopes of
the curves give $1/z=0.743(5)$ and $0.764(6)$ for $\Delta = 1.75$ and $2.5$ respectively.
The correction to scaling at early times is not universal, but usually in a power-law form.
After introducing an empirical power-law correction to scaling,
i.e., $F(t) \sim t^{1/z}(1+c/t)$, the fitting to the numerical data extends
to a larger interval, yield $1/z=0.758(6)$ and $0.774(6)$ respectively, and improve the results by about two percent.

To calculate the logarithmic derivative $\partial _\tau \ln v(t,\tau) =\partial
_\tau v(t,\tau)/v(t,\tau)$, we interpolate $v(t,\tau)$ between
different $H$, e.g., in the interval $[0.927, 0.943]$ for
$\Delta=1.75$, and $[0.946, 0.956]$ for $\Delta=2.5$. In
Fig.~\ref{Ft}(b), $\partial _\tau \ln v(t,\tau)$ is
plotted for $\Delta = 1.75$ and $2.5$ at $H_c$. A power-law behavior is observed for $t \geq 100$.
According to Eq.~(\ref{equ140}), from the slopes of
the curves one measures $1/\nu z=0.699(9)$ and $0.683(9)$ for $\Delta = 1.75$ and $2.5$ respectively.
With a power-law correction to scaling, the fitting to the numerical data extends to early times,
although the fitted exponents remain almost unchanged, $1/\nu z=0.703(9)$ and $0.689(9)$ respectively.

\begin{figure}[ht]
\epsfysize=7.1cm \epsfclipoff \fboxsep=0pt
\setlength{\unitlength}{1.cm}
\begin{picture}(10,7)(0,0)
\put(-3.2,-0.3){{\epsffile{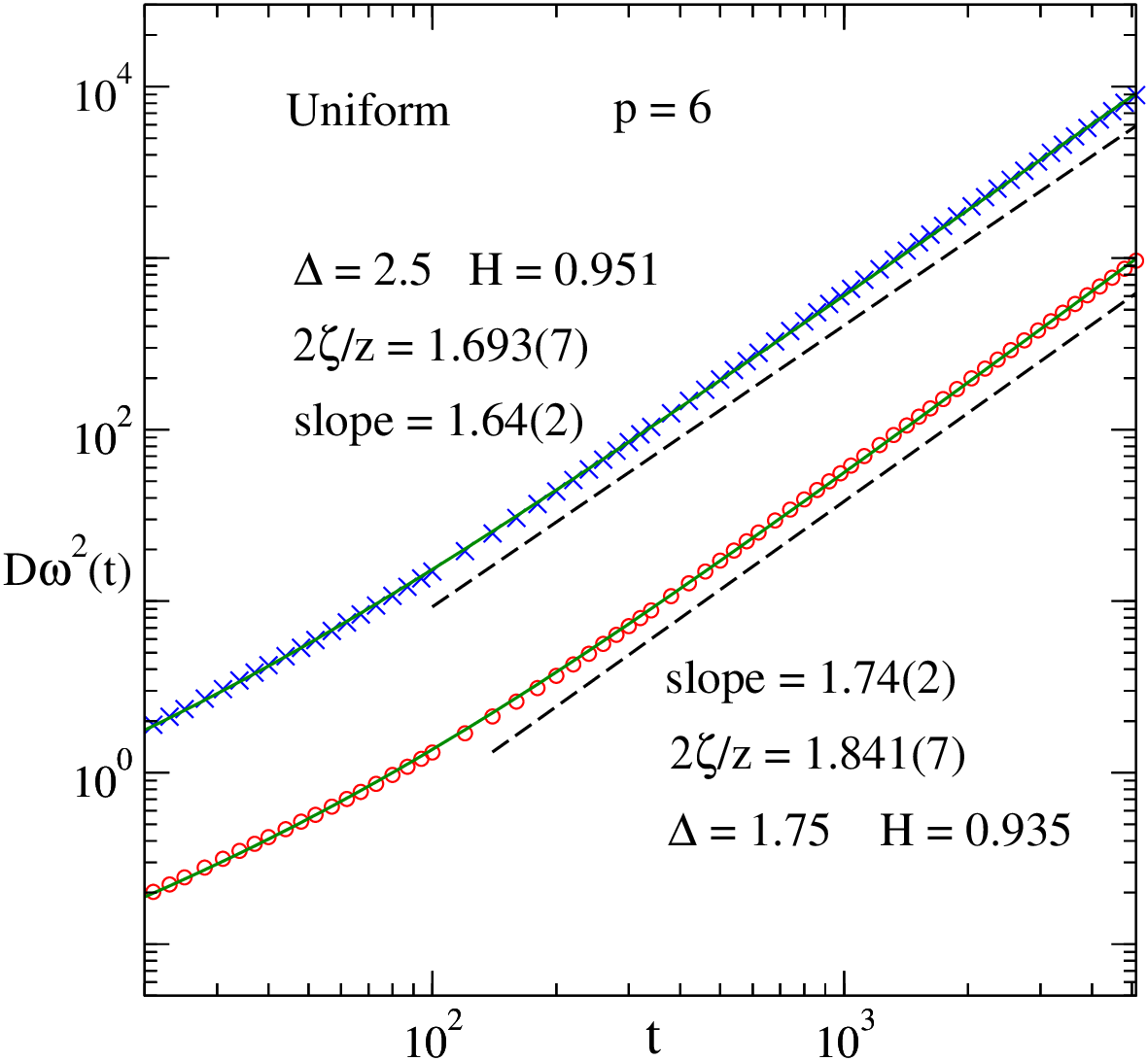}}}\epsfysize=7.1cm
\put(5.3,-0.3){{\epsffile{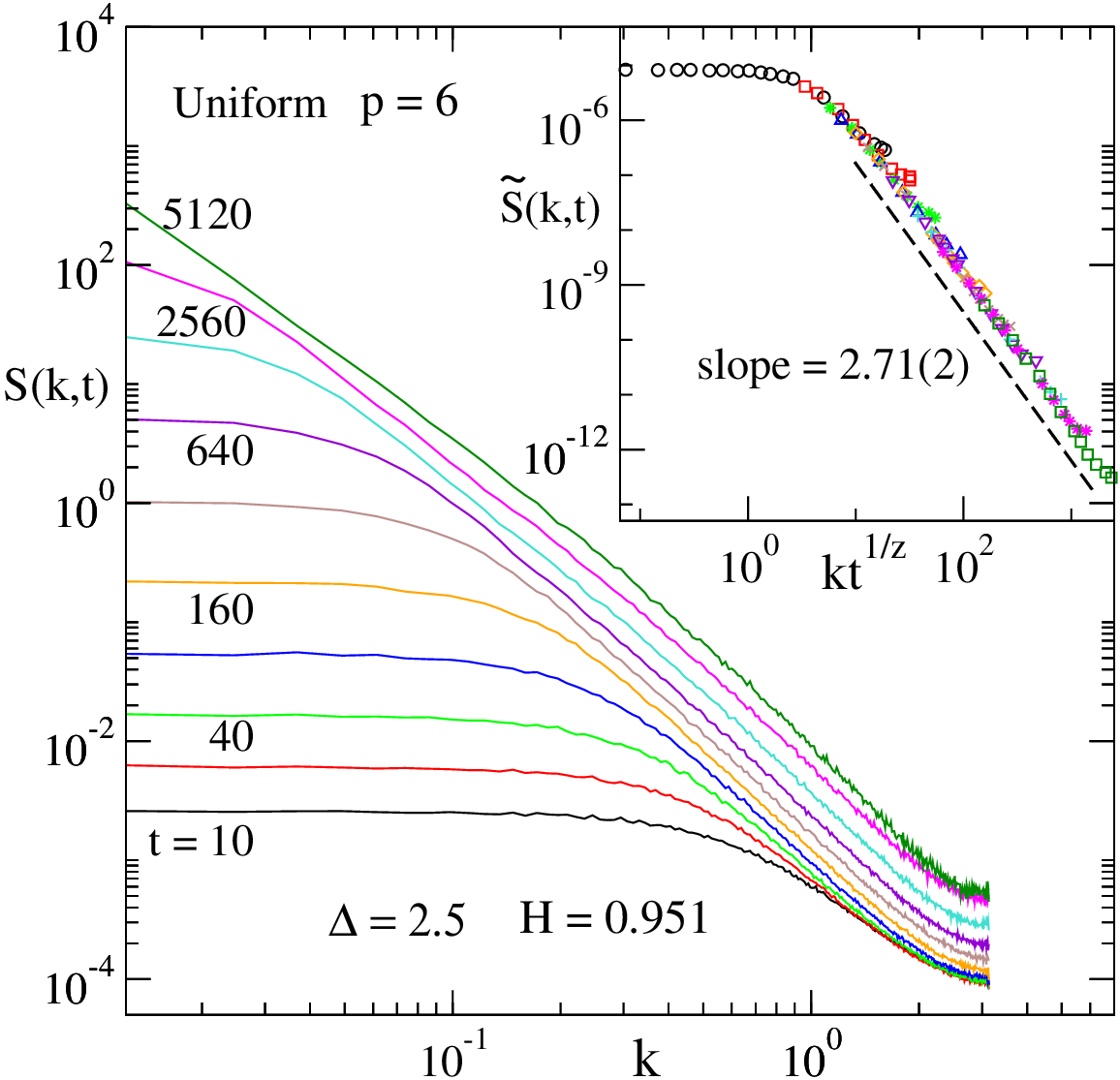}}}
\put(-2.4,-0.4){\footnotesize{(a)}}\put(6.3,-0.4){\footnotesize{(b)}}
\end{picture}
\caption{(Color online)(a) The pure roughness function $D\omega^2(t)$ is displayed
for $\Delta = 1.75$ and $2.5$ at $H_c$. Solid lines represent
power-law fits with corrections to scaling. For clarity,
the curve for $\Delta = 2.5$ is shifted up by a factor of five. (b) The structure factor
$S(k,t)$ is plotted for $\Delta=2.5$ at $H_c$. In the inset, data
collapse for $\tilde S(k,t) = S(k,t) t^{-(2 \zeta +1)/z}$ for different $t$ is shown.
In both (a) and (b), the results are for the uniform distribution of the random fields, and dashed lines show power-law fits.
 }\label{Dwt}
\end{figure}

In Fig.~\ref{Dwt}(a), the pure roughness function $D\omega^2(t)$ is
displayed for different strengths of the random fields at $H_c$.
A power-law behavior is observed for $t \geq 100$.
According to Eq.~(\ref{equ170}), a direct measurement from the slopes of
the curves gives $2 \zeta /
z=1.74(2)$ and $1.64(2)$ for $\Delta = 1.75$ and $2.5$ respectively.
With a power-law correction to scaling, the results are refined to be $2 \zeta /
z=1.841(7)$ and $1.693(7)$ respectively.

In Fig.~\ref{Dwt}(b), the structure factor $S(k,t)$ as a function of $k$ is displayed for $\Delta=2.5$ at
$H_c$. Obviously, for $u=kt^{1/z} \ll 1$, $u^{-(2\zeta+1)}f_s(u)$
approaches a constant. For $u \gg 1$,
$u^{-(2\zeta+1)}f_s(u)$ exhibits a power-law behavior. From the slope of the curve, one extracts
the exponent $2\zeta_s +1 =2.71(2)$,
i.e., the spectral roughness exponent $\zeta_s=0.86(1)$, by Eqs.~(\ref{equ190}) and (\ref{equ200}).
According to Eq.~(\ref{equ190}),
data collapse of $\tilde S(k,t) =S(k,t) t^{-(2 \zeta +1)/z}$ for different $t$ is shown in the inset. The
exponents used for the data collapse are $\zeta=1.09$ and $z=1.29$
obtained from Fig.~\ref{Ft}(a) and \ref{Dwt}(a) respectively.
Since $\zeta_s$ is smaller than $\zeta$, the high-$k$ mode of $S(k,t)$ grows with time by a power law,
as Eqs.~(\ref{equ190}) and (\ref{equ200}) indicate.
In other words, the anomalous dimension of the fluctuations in a small scale is different from that in a large scale.

\begin{figure}[ht]
\epsfysize=7.1cm \epsfclipoff \fboxsep=0pt
\setlength{\unitlength}{1.cm}
\begin{picture}(10,7)(0,0)
\put(-3.2,-0.3){{\epsffile{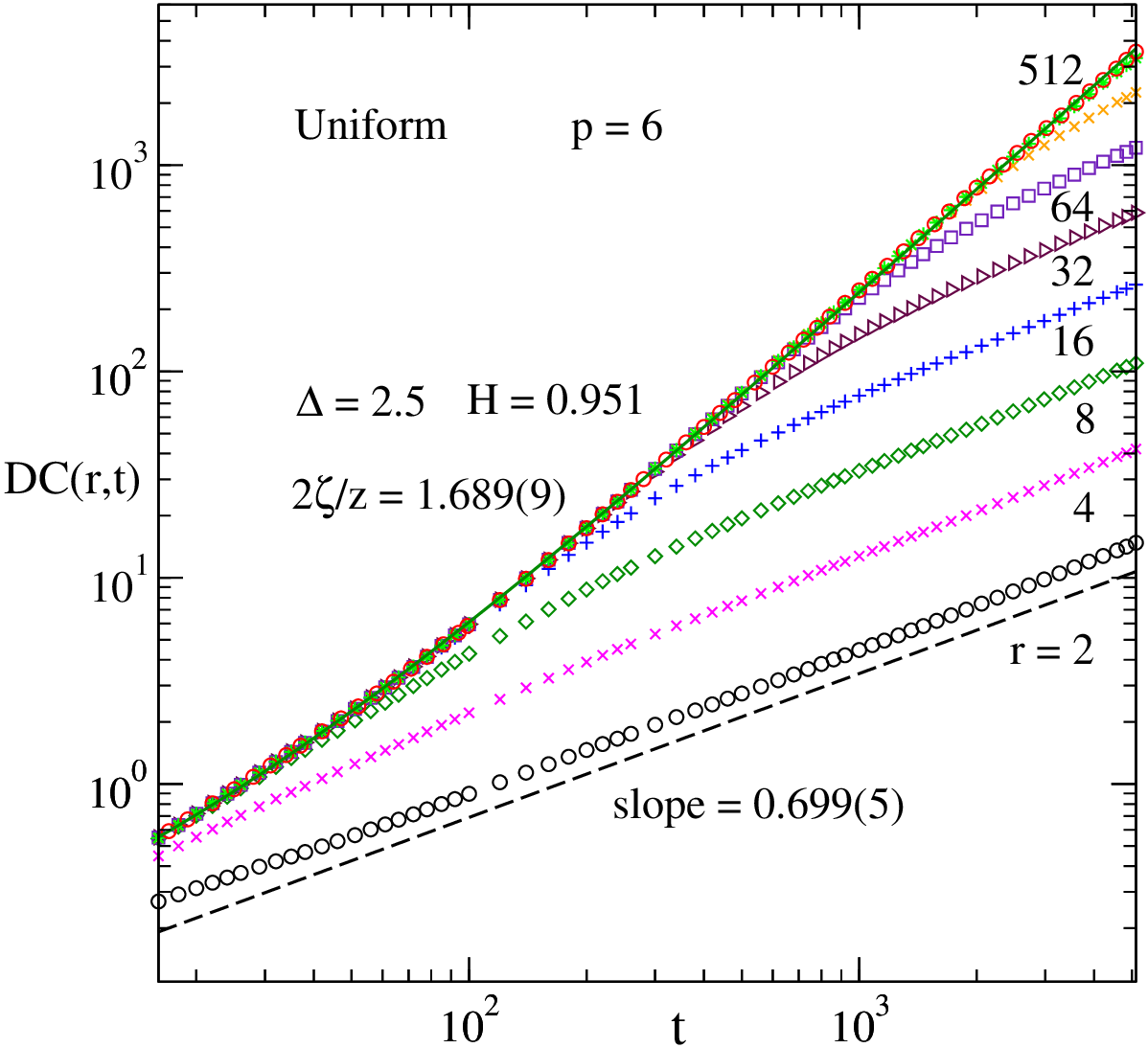}}}\epsfysize=7.1cm
\put(5.2,-0.3){{\epsffile{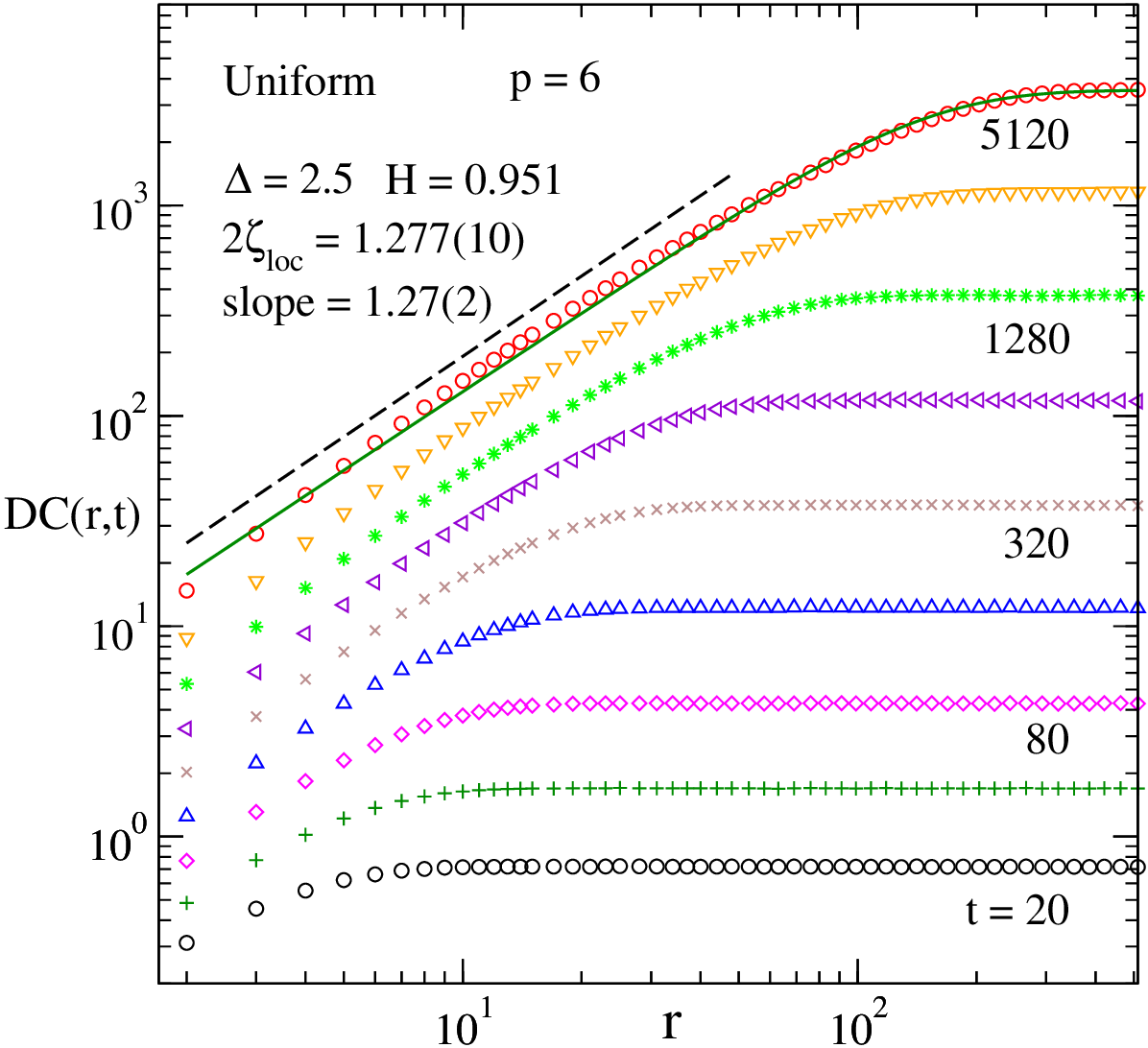}}}
\put(-2.4,-0.4){\footnotesize{(a)}}\put(6.3,-0.4){\footnotesize{(b)}}
\end{picture}
\caption{(Color online)(a) The pure height correlation function
$DC(r,t)$ is displayed for $\Delta=2.5$ at $H_c$. The solid line
shows a power-law fit with a correction to scaling.
(b) $DC(r,t)$ is plotted as a function of $r$
for $\Delta=2.5$ at $H_c$. The solid line shows a power-law fit with
a hyperbolic tangent correction. In both (a) and (b),
the results are for the uniform distribution of the random fields,
and dashed lines show power-law fits.}\label{dc}
\end{figure}

In Fig.~\ref{dc}(a), the pure height correlation function $DC(r,t)$
is displayed for $\Delta =2.5$ at $H_c$. For $r\gg \xi
(t)$, e.g., $r=512$, one extracts the exponent $2\zeta /z =1.689(9)$
by Eq.~(\ref{equ180}), consistent with that from Fig.~\ref{Dwt}(a).
For $r\ll \xi (t)$, e.g., $r=2$, $DC(r,t)$
should be independent of $t$ for a normal interface with $\zeta
=\zeta_{loc}$, according to Eq.~(\ref{equ180}). But $DC(r,t)$ of
$r=2$ in Fig.~\ref{dc}(a) clearly increases with time $t$. A
power-law behavior is observed. According to Eq.~(\ref{equ180}), the
slope of the curve gives $2(\zeta-\zeta_{loc})/z =0.699(5)$. From
the measurements of $\zeta$ and $z$ from Fig.~\ref{Ft}(a) and
\ref{Dwt}(a), we calculate the local roughness exponent
$\zeta_{loc}=0.64(1)$.
Since $\zeta_{loc}$ is smaller than $\zeta$, $DC(r,t)$ for a small $r$ grows with time by a power law,
according to Eq.~(\ref{equ180}). This behavior is in accord with that of the high-$k$ mode of $S(k,t)$.
We empirically observe $\zeta-\zeta_{loc} \approx 2(\zeta-\zeta_s)$, although we are not able to theoretically derive it.

In Fig.~\ref{dc}(b), the pure height correlation function $DC(r,t)$
is plotted as a function of $r$.
For $r\ll \xi(t)$, there exists a power-law behavior, but this regime is rather
limited. From the slope of
the curves one estimates $2\zeta_{loc}=1.27(2)$, i.e.,
$\zeta_{loc} =0.64(1)$.
For a large $r$, there emerge strong corrections to scaling. With a specific form of corrections to scaling \cite{jos96},
\begin{equation}
DC(r) \sim [\tanh (r/c) ]^{2\zeta_{loc}}, \label{equ230}
\end{equation}
the fitting to the numerical
extends to a larger interval, and yields $\zeta_{loc} =0.639(5)$, in agreement with that from the slope of the curves.
This result is also consistent with that from Fig.~\ref{dc}(a) within errors.

From the measurements of $\beta /\nu z$, $1/z$, $1/\nu z$, $2 \zeta
/ z$, $2\zeta_s +1$, $2(\zeta-\zeta_{loc})/z$ and $2 \zeta_{loc}$ in
Fig.~\ref{vt} to \ref{dc} respectively, we calculate the static
critical exponents $\beta$ and $\nu$, the dynamic critical exponent
$z$, the roughness exponents $\zeta, \zeta_{loc}$ and $\zeta_s$ for the $6$-state DRFCM model. All
the results are summarized in Table~\ref{t1}, compared with
those of the QEW equation. Obviously, the critical exponents vary with the
strength of the random fields.
The critical exponent $\nu$ increases, and all other critical exponents
decrease with the strength of the random fields, staring from $\Delta=1.25$.
In general, the critical exponents of the DRFCM model differ from those of the QEW equation
by $10$ to $30$ percent. In our opinion, this difference is induced by overhangs and islands, and also intermediate regimes created by the intermediate states of the spin $\vec S_i$ \cite{zho09,zho10}.

At this stage, the dependence of the critical exponents on the strength of the random fields for the $6$-state DRFCM model
is not much different from that for the case of $p=2$, i.e., the DRFIM model \cite{qin12}, although the microscopic structure of
the domain interface including overhangs and islands, and intermediate regimes, becomes more complicated.
This might be related to the definition of the height function in Eq~(\ref{equ30}) which averages over the orientations of
the spins.

For the roughening interfaces, according to Ref.~\cite{ram00}, the roughness exponents $\zeta,
\zeta_{loc}$ and $\zeta_s$ are not always independent, and there are four
different subclasses, namely, Family-Vicsek, super-rough, intrinsic
and faceted. In Ref.~\cite{che10}, however, it suggests that there is a new
subclass of anomalous roughening, with $\zeta \neq
\zeta_{loc} \neq \zeta_s$ and $\zeta_{loc} \neq 1$.
Our results in Table~\ref{t1} indicate that the growth dynamics of the domain interface in the DRFCM model,
similar to that in the DRFIM model \cite{qin12}, belongs to such a subclass.

Looking at Fig.~\ref{evolution}, one might catch such a scenario that the domain-wall motion could be decomposed into modes:
one is the growth of the width of the domain wall, and another is the roughening of the domain wall.
The subtle point here is how to define the width of the domain wall. Essentially, the width of the domain wall should be generated by overhangs and islands,
and intermediate regimes. Therefore, one possibility is to define the width of the domain wall by the size
of overhangs and islands, and intermediate regimes. For $p=2$, there are no intermediate regimes, and the dynamic relaxation of overhangs and islands have been carefully analyzed
in Ref.~\cite{zho10}. The average size of overhangs and islands does grow with time by a power law. Denoting the velocity of overhangs and islands
by $\delta v$, and the velocity of the domain wall defined by the magnetization by $v_M$, one obtains $\delta v/ v_M \sim  t^\theta$, with $\theta \approx 0.5$ \cite{zho10}.
For p=6, one may similarly define the width of the domain wall, although due to the existence of intermediate regimes, the fluctuation is somewhat larger. The exponent $\theta$ is estimated to be $\theta \approx 0.25$.
Since the paper is already lengthy, details of these contents will not be presented here.

However, it is still not possible to find a decomposition scheme such that the roughening mode of the domain wall is the same as that of the QEW equation. In some sense, the height function defined by the magnetization already minimizes the effect of overhangs and islands, and intermediate state, but its dynamic evolution is still affected by overhangs and islands, an intermediate states.

\subsection{Gaussian distribution of random fields}

We now consider the Gaussian distribution of the random fields.
As shown in Fig.~\ref{Hc} and Tables~\ref{t3} and \ref{t2}, the transition field $H_c$ varies
also with the state number $p$ of the spin $\vec S_i$ and the standard deviation $\sigma$ of the Gaussian distribution.
Since the Gaussian distribution is unbounded, the
crossover of the depinning transition from the second order to the first order
is not present. Qualitatively, the functions $H_c(p)$ and $H_c(\sigma)$ for the Gaussian distribution
are similar to those for the uniform distribution. In particular, if one makes a correspondence $\Delta \to 1.5 \sigma$,
the transition field $H_c$ for the Gaussian distribution looks rather close to that for the uniform distribution,
as shown in Fig.~\ref{Hc} and Tables~\ref{t3} and \ref{t2}.
Quantitatively, $H_c(\sigma)$ for the Gaussian distribution fits nicely to a parabolic curve
as can be seen in Fig.~\ref{Hc}(b).

The critical exponents for the $6$-state DRFCM model with the Gaussian distribution of the random fields are given in Table~\ref{t2}.
Although the dependence of the critical exponents on the strength of the random fields looks similar for both
the Gaussian and uniform distributions in the regimes $\Delta \geq 1.5$ and $ \sigma \geq 1$,
the critical exponents do vary with the form and strength of the random fields. The strong
universality is violated. For $p=2$, i.e., the DRFIM model, some critical exponents or their combinations,
e.g., $\beta$ and $(2\zeta+1)/z$, are 'quasi universal'. In other words, these critical exponents
are independent of the form and strength of the random fields in certain regimes, e.g., $\Delta \geq 1.5$ and $ \sigma \geq 1$.
However, such a quasi universality does not exist for the $6$-state DRFCM model.
This should be due to the multiple orientations of the spin $\vec S_i$.
The dynamic evolution of the domain interface is more sensitive to the quenched disorder.

\subsection{Orientations of ordered domains}

For $p=2$, i.e., the DRFIM model, there are only two possible orientations of the magnetization of the two ordered domains,
and the domain interface is unique. For $p \geq 4$, different orientations of the two ordered domains
may lead to different domain interfaces. For $p=3$, there exist two domain interfaces, but
they are equivalent. Now, an interesting question is, whether the depinning phase transitions of the domain interfaces
with different orientations of the two ordered domains
occur at a same transition field, and belong to a same universality class, or not.

\begin{figure}[ht]
\epsfysize=7.1cm \epsfclipoff \fboxsep=0pt
\setlength{\unitlength}{1.cm}
\begin{picture}(10,7)(0,0)
\put(-3.2,-0.3){{\epsffile{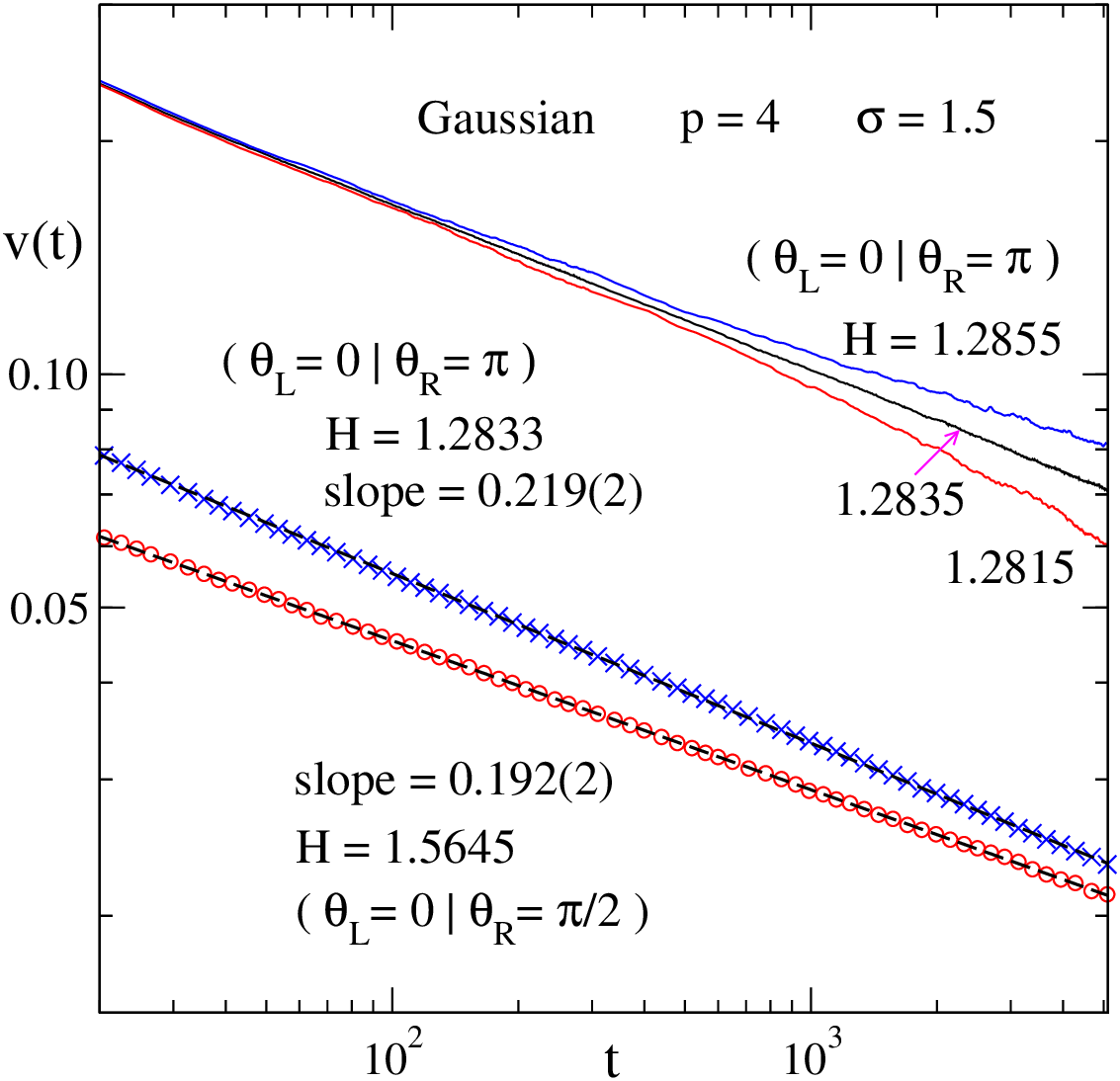}}}\epsfysize=7.1cm
\put(5.2,-0.3){{\epsffile{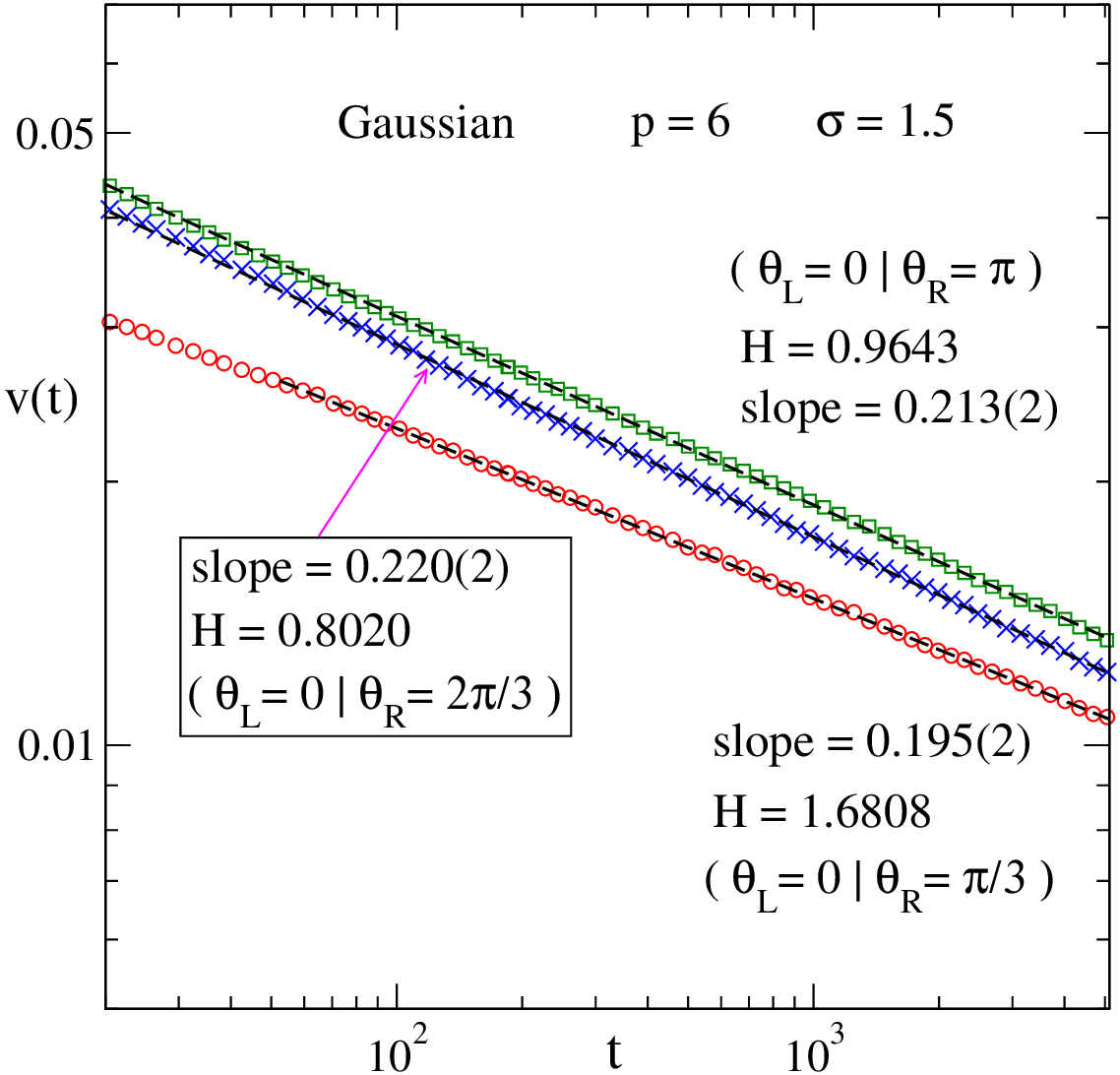}}}
\put(-2.4,-0.4){\footnotesize{(a)}}\put(6.3,-0.4){\footnotesize{(b)}}
\end{picture}
\caption{(Color online)(a) The interface velocity $v(t)$ is plotted for the $4$-state DRFCM model with the Gaussian distribution of the random fields
at different driving field $H$ with solid lines, for the initial orientation $\theta_R=\pi$. For comparison,
$v(t)$ is also shown for $\theta_R=\pi$ and $\pi/2$ at $H_c$ respectively with crosses and circles.
For clarity, the solid lines are shifted up by a factor of three.
(b) $v(t)$ is displayed for the $6$-state DRFCM model at $H_c$,
with different $\theta_R$. In both (a) and (b), dashed lines represent
power-law fits. }\label{p4p6}
\end{figure}

To determine the transition field $H_c$, we again make use of Eq.~(\ref{equ120}).
In Fig.~\ref{p4p6}(a), one locates $H_c$ for the $4$-state DRFCM model with the Gaussian distribution of the random fields at $\sigma=1.5$,
by searching for the best power-law behavior of the interface velocity $v(t,\tau)$.
Obviously, the transition field $H_c$ depends on the orientations of the two ordered domains.
For $\theta_L=0$ and $\theta_R=\pi$, $H_c=1.2833(3)$; for $\theta_L=0$ and $\theta_R=\pi/2$, $H_c=1.5645(3)$.
In Table~\ref{t4}, the results for $p=4$, $6$ and $8$ are summarized.

Since the driving field $H$ is applied in the $x$ direction, one may naively think that the smaller $\theta_R$ is,
the easier it is to drive the domain wall to propagate. Therefore, the transition field $H_c$ may be expected to decrease with $\theta_R$.
Looking at Table~\ref{t4}, however, one finds rather unusual dependence of $H_c$ on $\theta_R$. As $\theta_R$ decreases, except for $\theta_R=\pi$,
$H_c$ does not decreases, rather increases! For $p=8$, for example, $H_c=0.6018(5)$ for $\theta_R=3\pi/4$,
while $H_c=1.7825(5)$ for $\theta_R=\pi/4$. The difference is very prominent.
It is much more difficult to drive the domain wall of $\theta_R=\pi/4$ to propagate.

Why does such an unusual phenomenon occur?
For $\theta_R \neq \pi$, due to the action of the driving field $H$ and the average field of the nearest neighbor spins $\vec S_j$,
a spin $\vec S_i$ on the domain interface rotates almost exclusively to the direction $\theta_i < \theta_R$.
In Fig.~\ref{evolution}(b), for example, the snapshot for $\theta_R=2\pi/3$ is displayed for the $6$-state DRFCM model.
Almost no spins orient in the directions $\theta_i=\pi$ and $4\pi/3$, and only a few spins in the direction $\theta_i=5\pi/3$.
In other words, for $\theta_R=2\pi/3$, spins may be flipped to $\theta_i=\pi/3$ and $0$, while for $\theta_R=\pi/3$, to $\theta_i=0$ only.
For $\theta_R=\pi/3$, $\theta_i$ is pinned if the flip to $\theta_i=0$ is forbidden by the random fields.
For $\theta_R=2\pi/3$, however, there is still another channel to $\theta_i=\pi/3$.
Therefore, $H_c$ of the domain interface with $\theta_R=\pi/3$ is larger.

The transition field $H_c$ for $\theta_R = \pi $ is smaller than that for the smallest $\theta_R$, while larger than others.
One may understand the former phenomenon following the argument above, but the latter is puzzling.
A possible reason might be that in the case of $\theta_R = \pi $, the average field of the nearest neighbor spins $\vec S_j$ in the domain interface is zero,
only the external field drives the spins to the $x$ direction, therefore, $H_c$ is somewhat larger.

Another issue one still needs to address is that according to Table~\ref{t4}, the transition field $H_c$ for the smallest $\theta_R$ increases with $p$.
This seems contradict with the fact that the depinning transition is absent in the limit $p \to \infty$.
Therefore we compute $H_c$ for the smallest $\theta_R$ up to $p=24$, and the results show that $H_c$ starts dropping around $p=16$.

With the procedure in Subsec. III A, we may determine all the critical exponents for the domain interfaces with different orientations of the two ordered domains.
In Fig.~\ref{p4p6}(a) and (b), for example, one measures the exponent $\beta/\nu z$ from the slopes of the curves $v(t)$ at $H_c$.
From the results summarized in Table~\ref{t4}, we observe that the critical exponents do change with the orientations of the two ordered domains,
and this phenomenon tends to be more prominent for larger $p$.
In other words, the depinning phase transitions of the domain interfaces with different orientations of the two ordered domains
belong to different universality classes.
As $\theta_R$ decreases, the dynamic exponent $z$, roughness exponents $\zeta$, $\zeta_{loc}$ and $\zeta_s$ decrease monotonously,
while the static exponent $\nu$ increases. The dependence of the static exponent $\beta$ on $\theta_R$ is somewhat complicated.
Roughly speaking, a larger $H_c$ corresponds to a smaller $\beta$.
The exponent $\beta/\nu$ is, however, almost independent of $\theta_R$, except for the smallest $\theta_R$.

Comparing the results in Table~\ref{t4} with those in Tables~\ref{t1} and \ref{t2}, one may find that as $\theta_R$ decreases, the dynamic effect of
the random fields tends to be stronger. In other words, for smaller $\theta_R$, $\theta_i$ is more sensitive to the random fields.
In any case, the domain interfaces always belong to the new subclass of anomalous roughening, with $\zeta \neq
\zeta_{loc} \neq \zeta_s$ and $\zeta_{loc} \neq 1$.

\section{Conclusion}

With Monte Carlo simulations, we have investigated the depinning phase transition
of the domain interface in the $2D$ $p$-state DRFCM
model, emphasizing the dynamic effect of the orientations of the magnetization of the two ordered domains. Based on the short-time dynamic approach, the
transition field and all the critical exponents are determined.
Numerical results are mainly presented for $p=6$, and also extended from $p=2$ to $12$.

The important results are summarized below:

i) For the uniform distribution of the random fields, there exists a critical strength $\Delta_c$ of the random fields,
separating the first-order depinning transition ($\Delta < \Delta_c$) and second-order one ($\Delta > \Delta_c$).
For the Gaussian distribution, the depinning transition is of second order.
For a fixed strength $\Delta$ of the random fields, the transition field $H_c \to 0$ in the limit $p \to \infty$.

ii) For the second-order depinning transition, the critical exponents do vary with
the form of the random-field distribution and the strength of the
random fields, and the strong universality is violated.
Further, we observe $\zeta \neq \zeta_{loc}\neq \zeta_s$, and $\zeta_{loc}<1$, and
it indicates that the roughening dynamics of the domain interface belongs to the new subclass suggested in Ref.~\cite{che10}.

iii) More importantly, the transition field $H_c$ and critical exponents change with the initial orientations of the magnetization of the two ordered domains.
$H_c$ exhibits rather unusual dependence on the orientation $\theta_R$ as shown in Table~\ref{t4}.
Unexpectedly, the dynamic effect of the random fields is the most significant for the domain interface
with the smallest $\theta_R$.

{\bf Acknowledgement} This work was supported in part by NNSF of China under Grant
Nos. 10875102 and 11075137.

%\bibliography{domain,zheng}

\begin{thebibliography}{10}

\bibitem{edw82}
{S.F. Edwards, D.R. Wilkinson}, Proc. R. Soc. London, Ser. A {\bf 381},  17
  (1982).

\bibitem{nat92}
{T. Nattermann, S. Stepanow, L.H. Tang, and H. Leschhorn}, J. Phys. II France
  {\bf 2},  1483  (1992).

\bibitem{ros03}
{A. Rosso, A.K. Hartmann, and W. Krauth}, Phys. Rev. {\bf E67},  021602
  (2003).

\bibitem{due05}
{O. Duemmer and W. Krauth}, Phys. Rev. {\bf E71},  061601  (2005).

\bibitem{kol06}
{A.B. Kolton, A. Rosso, E.V. Albano, and T. Giamarchi}, Phys. Rev. {\bf B74},
  140201(R)  (2006).

\bibitem{bus08}
{S. Bustingorry, A.B. Kolton, and T. Giamarchi}, Europhys. Lett. {\bf 81},
  26005  (2008).

\bibitem{now98}
{U. Nowak and K. D. Usadel}, Europhys. Lett. {\bf 44},  634  (1998).

\bibitem{rot99}
{L. Roters, A. Hucht, S. L\"ubeck, U. Nowak, and K.D. Usadel}, Phys. Rev. {\bf
  E60},  5202  (1999).

\bibitem{rot01}
{L. Roters, S. L\"ubeck, and K. D. Usadel}, Phys. Rev. {\bf E63},  026113
  (2001).

\bibitem{zho09}
{N.J. Zhou, B. Zheng, and Y.Y. He}, Phys. Rev. B {\bf 80},  134425  (2009).

\bibitem{zho10}
{N.J. Zhou and B. Zheng}, Phys. Rev. E {\bf 82},  031139  (2010).

\bibitem{don12}
{R.H. Dong, B. Zheng and N.J. Zhou}, EPL {\bf 98},  36002  (2012).

\bibitem{met07}
{P.J. Metaxas, J.P. Jamet, A. Mougin, M. Cormier, J. Ferr\'e, V. Baltz, B.
  Rodmacq, B. Dieny, and R. L. Stamps}, Phys. Rev. Lett. {\bf 99},  217208
  (2007).

\bibitem{lem98}
{S. Lemerle, J. Ferr\'{e}, C. Chappert, V. Mathet, T. Giamarchi, and P. Le
  Doussal}, Phys. Rev. Lett. {\bf 80},  849  (1998).

\bibitem{jos98}
{M. Jost, J. Heimel and T. Kleinefeld}, Phys. Rev. {\bf B57},  5316  (1998).

\bibitem{zho10a}
{N.J. Zhou, B. Zheng and D.P. Landau}, EPL {\bf 92},  36001  (2010).

\bibitem{ros01}
{A. Rosso and W. Krauth}, Phys. Rev. Lett. {\bf 87},  187002  (2001).

\bibitem{zhe98}
B. Zheng, Int. J. Mod. Phys. {\bf B12},  1419  (1998), review article.

\bibitem{luo98}
{H.J. Luo, L. Sch\"ulke, and B. Zheng}, Phys. Rev. Lett. {\bf 81},  180
  (1998).

\bibitem{rod07}
{G. Rodr¨ªguez-Rodr\'{i}guez, A.P. Junquera, M. V\'{e}lez, J.V. Anguita, J.I.
  Mart\'{i}n, H. Rubio, and J.M. Alameda}, J. Phys. D: Appl. Phys. {\bf 40},
  3051  (2007).

\bibitem{kol09}
{A.B. Kolton, G. Schehr, and P.Le Doussal}, Phys. Rev. Lett. {\bf 103},  160602
   (2009).

\bibitem{cat12}
{G. Catalan, J. Seidel, R. Ramesh, and J.F. Scott}, Rev. Mod. Phys. {\bf 84},
  119  (2012).

\bibitem{rhe10}
{J. Rhensius, L. Heyne, D. Backes, S. Krzyk, L. J. Heyderman, L. Joly, F.
  Nolting, and M. Kl\"aui}, Phys. Rev. Lett. {\bf 104},  067201  (2010).

\bibitem{uhl10}
{V. Uhl\'{i}\u{r}, S. Pizzini, N. Rougemaille, J. Novotn\'{y}, V. Cros, E.
  Jim\'{e}nez, G. Faini, L. Heyne, F. Sirotti, C. Tieg, A. Bendounan, F.
  Maccherozzi, R. Belkhou, J. Grollier, A. Anane, and J. Vogel}, Phys. Rev. B
  {\bf 81},  224418  (2010).

\bibitem{gou10}
{A. Goussev, J.M. Robbins, and V. Slastikov}, Phys. Rev. Lett. {\bf 104},
  147202  (2010).

\bibitem{mar07}
{E. Martinez, L. Lopez-Diaz, L. Torres, C. Tristan, and O. Alejos}, Phys. Rev.
  B {\bf 75},  174409  (2007).

\bibitem{shi07}
{Y.H. Shin, I. Grinberg, I.W. Chen, and A.M. Rappe}, Nature {\bf 449},  881
  (2007).

\bibitem{luo07}
{M.B. Luo and X. Hu}, Phys. Rev. Lett. {\bf 98},  267002  (2007).

\bibitem{nel11}
{C.T. Nelson, B. Winchester, Y. Zhang, S.J. Kim, A. Melville, C. Adamo, C.M.
  Folkman, S.H. Baek, C.B. Eom, D.G. Schlom, L.Q. Chen, and X.Q. Pan}, Nano
  Lett. {\bf 11},  828  (2011).

\bibitem{jia11}
{C.L. Jia, K.W. Urban, M. Alexe, D. Hesse, and I. Vrejoiu}, Science {\bf 331},
  1420  (2011).

\bibitem{cha86}
{M.S.S. Challa and D.P. Landau}, Phys. Rev. B {\bf 33},  437  (1986).

\bibitem{cze96}
{P. Czerner and U. Ritschel}, Phys. Rev. E {\bf 53},  3333  (1996).

\bibitem{bae09}
{S.K. Baek, P. Minnhagen, H. Shima, and B.J. Kim}, Phys. Rev. E {\bf 80},
  011133  (2009).

\bibitem{lap06}
{C.M. Lapilli, P. Pfeifer and C. Wexler}, Phys. Rev. Lett. {\bf 96},  140603
  (2006).

\bibitem{bri07}
{A.F. Brito, J.A. Redinz and J.A. Plascak}, Phys. Rev. E {\bf 75},  046106
  (2007).

\bibitem{bri10}
{A.F. Brito, J.A. Redinz, and J.A. Plascak}, Phys. Rev. E {\bf 81},  031130
  (2010).

\bibitem{ji91}
{H. Ji and M.O. Robbins}, Phys. Rev. A {\bf 44},  2538  (1991).

\bibitem{koi10}
{B. Koiller and M.O. Robbins}, Phys. Rev. B {\bf 82},  064202  (2010).

\bibitem{dro98}
{B. Drossel and K. Dahmen}, Eur. Phys. J. B {\bf 3},  485  (1998).

\bibitem{ram00}
{J.J. Ramasco, J.M. L\'{o}pez and M.A. Rodr\'{\i}guez}, Phys. Rev. Lett. {\bf
  84},  2199  (2000).

\bibitem{che10}
{Y.J. Chen, Y. Nagamine, T. Yamaguchi and K. Yoshikawa}, Phys. Rev. E {\bf 82},
   021604  (2010).

\bibitem{jan89}
{H.K. Janssen, B. Schaub, and B. Schmittmann}, Z. Phys. {\bf B73},  539
  (1989).

\bibitem{jos96}
{M. Jost and K.D. Usadel}, Phys. Rev. {\bf B54},  9314  (1996).

\bibitem{zho08}
{N.J. Zhou and B. Zheng}, Phys. Rev. E {\bf 77},  051104  (2008).

\bibitem{bak08}
{B. Bak\'o, D. Weygand, M. Samaras, W. Hoffelner, and M. Zaiser}, Phys. Rev.
  {\bf B78},  144104  (2008).

\bibitem{qin12}
{X.P. Qin, B. Zheng and N.J. Zhou}, J. Phys. A: Math. Theor. {\bf 45},  115001
  (2012).

\end{thebibliography}
%\bibliographystyle{prsty}

\begin{table}[h]
\begin{tabular}[t]{ c |c| c c || c| c|c c}
\hline\hline
  p &  $\Delta_c$  &  $H_c$       &          & p  &  $\Delta_c$ &  $H_c$   &   \\
\hline
   &     & $\Delta=1.5$ & $\sigma=1.0$ &     &    & $\Delta=1.5$ & $\sigma=1.0$\\
\hline
 2  & 1.000(1) & 1.2933(2) & 1.4205(3) &  8  & 0.42(1) & 0.757(1)  & 0.778(2)  \\
 3  & 1.000(1) & 1.341(1)  & 1.513(2)  &  9  & 0.53(1) & 0.594(2)  & 0.607(2)  \\
 4  & 1.000(1) & 1.192(1)  & 1.255(1)  &  10 & 0.41(1) & 0.634(2)  & 0.664(2)  \\
 5  & 0.625(5) & 0.906(1)  & 0.968(2)  &  11 &         & 0.515(2)  & 0.526(2)  \\
 6  & 0.51(1)  & 0.921(1)  & 0.946(1)  &  12 &         & 0.553(3)  & 0.585(3)  \\
 7  & 0.58(1)  & 0.705(1)  & 0.727(2)  &  36 &         & $<$~0.001 & $<$~0.001 \\
\hline\hline
\end{tabular}
\caption{The critical strength $\Delta_c$ for the uniform distribution of the random fields,
and the depinning transition field $H_c$ at
$\Delta=1.5$ and $\sigma=1.0$ for the uniform and Gaussian
distributions of the random fields respectively.} \label{t3}
\end{table}

\begin{table}[h]
\begin{tabular}[t]{ c | c c c c | c | c| c | c }
\hline \hline
                 &     &    $v(t)$      &      &   & $F(t)$,$C(r,t)$  & $\omega^2(t)$,$C(r,t)$ & $C(r,t)$      & $S(k,t)$ \\
\hline

$\Delta$         &$H_c$     &$\beta$   & $\nu$   &$\beta/\nu$  & $z$     & $\zeta$  & $\zeta_{loc}$  & $\zeta_s$   \\
\hline
1.25             & 0.913(1) & 0.491(8) & 0.95(2) & 0.517(4) & 1.40(1) & 1.24(1)  & 0.78(1)  & 1.05(1)   \\
1.5              & 0.921(1) & 0.415(8) & 1.04(2) & 0.399(4) & 1.37(1) & 1.22(1)  & 0.75(1)  & 1.00(1)   \\
1.75             & 0.935(1) & 0.358(6) & 1.06(2) & 0.338(3) & 1.33(1) & 1.21(1)  & 0.72(1)  & 0.97(1)   \\
2.0              & 0.954(1) & 0.333(6) & 1.08(2) & 0.308(3) & 1.32(1) & 1.18(1)  & 0.70(1)  & 0.92(1)   \\
2.25             & 0.963(1) & 0.313(6) & 1.10(2) & 0.285(3) & 1.31(1) & 1.13(1)  & 0.67(1)  & 0.89(1)   \\
2.5              & 0.951(1) & 0.302(6) & 1.12(2) & 0.270(3) & 1.29(1) & 1.09(1)  & 0.64(1)  & 0.86(1)   \\

\hline
DRFIM           & 1.2933(2) & 0.295(6) & 1.02(2) & 0.289(3) & 1.33(1) & 1.14(1)  & 0.74(1)  & 0.94(1)   \\
\hline
QEW             &           & 0.33(2)  & 1.33(4) & 0.25(2)   & 1.50(3) & 1.25(1)  & 0.98(6)  & 1.25(1)   \\
\hline \hline
\end{tabular}
\caption{The depinning transition field and critical exponents of
the $6$-state DRFCM model for different strengths of the random
fields with the uniform distribution are compared with those of the
QEW equation~\cite{due05,kol06,bus08} and DRFIM model with
$\Delta=1.5$ \cite{zho09,zho10}.} \label{t1}
\end{table}

\begin{table}[h]
\begin{tabular}[t]{ c | c c c c | c | c| c | c }
\hline \hline
                 &     &    $v(t)$      &        &  & $F(t)$,$C(r,t)$  & $\omega^2(t)$,$C(r,t)$ & $C(r,t)$      & $S(k,t)$ \\
\hline

$\sigma$         &$H_c$      &$\beta$   & $\nu$  &$\beta/\nu$  & $z$     & $\zeta$  & $\zeta_{loc}$  & $\zeta_s$   \\
\hline
1.0              & 0.946(1) & 0.380(6) & 1.00(2) & 0.380(3) & 1.35(1) & 1.26(1)  & 0.76(1)  & 0.97(1)   \\
1.2              & 0.965(1) & 0.337(6) & 1.05(2) & 0.321(3) & 1.33(1) & 1.19(1)  & 0.73(1)  & 0.90(1)   \\
1.4              & 0.968(1) & 0.321(6) & 1.13(2) & 0.284(3) & 1.31(1) & 1.12(1)  & 0.69(1)  & 0.83(1)   \\
1.6              & 0.957(1) & 0.312(6) & 1.18(2) & 0.264(3) & 1.29(1) & 1.06(1)  & 0.65(1)  & 0.80(1)   \\
\hline \hline
\end{tabular}
\caption{The depinning transition field and critical exponents of
the $6$-state DRFCM model for different strengths of the random
fields with the Gaussian distribution. } \label{t2}
\end{table}

\begin{table}[h]
\begin{tabular}[t]{ l| l l | c c c c | c | c | c | c  }
\hline \hline
 &         &       &     &    $v(t)$      &       &  & $F(t)$,$C(r,t)$  & $\omega^2(t)$,$C(r,t)$ & $C(r,t)$ & $S(k,t)$   \\
\hline
 &  $\theta_L$  &   $\theta_R$    &$H_c$      &$\beta$   & $\nu$ &$\beta/\nu$  & $z$     & $\zeta$  & $\zeta_{loc}$ & $\zeta_s$    \\
\hline
$p=4$ &  0 & $\pi $         & 1.2833(3)  & 0.319(6) & 1.10(2) & 0.290(3) & 1.32(1) & 1.11(1)  & 0.63(1) & 0.81(1)   \\
 & 0 & $ \frac{1}{2}\pi $   & 1.5645(3)  & 0.280(6) & 1.16(2) & 0.241(3) & 1.26(1) & 1.05(1)  & 0.60(1) & 0.76(1)   \\
\hline \hline

$p=6$ &  0 & $\pi $         & 0.9643(5)  & 0.319(6) & 1.15(2) & 0.277(3) & 1.30(1) & 1.10(1)  & 0.66(1)  & 0.81(1)  \\
 & 0 &  $ \frac{2}{3}\pi $  & 0.8020(5)  & 0.325(6) & 1.17(2) & 0.278(3) & 1.26(1) & 1.03(1)  & 0.57(1)  & 0.72(1)  \\
 & 0 &  $\frac{1}{3}\pi $   & 1.6808(5)  & 0.287(6) & 1.18(2) & 0.243(3) & 1.25(1) & 1.02(1)  & 0.56(1)  & 0.67(1)  \\

\hline\hline
$p=8$ & 0 &  $ \pi $         & 0.7841(5)  & 0.314(6) & 1.15(2) & 0.273(3) & 1.39(1) & 1.15(1)  & 0.65(1) & 0.90(1)   \\
 &  0 & $\frac{3}{4}\pi $    & 0.6018(5)  & 0.322(6) & 1.16(2) & 0.278(3) & 1.34(1) & 1.08(1)  & 0.57(1) & 0.83(1)   \\
 &  0 & $ \frac{1}{2}\pi $   & 0.6895(5)  & 0.334(6) & 1.21(2) & 0.276(3) & 1.26(1) & 1.01(1)  & 0.54(1) & 0.65(1)   \\
 &  0 & $\frac{1}{4}\pi $    & 1.7825(5)  & 0.308(6) & 1.25(2) & 0.246(3) & 1.21(1) & 0.98(1)  & 0.52(1) & 0.63(1)   \\
\hline \hline

\end{tabular}
\caption{The depinning transition field and critical exponents of
the $p$-state clock model with the Gaussian distribution at $\sigma
= 1.5 $ for different initial orientations of the magnetization of
the two ordered domains. } \label{t4}
\end{table}

\end{document}